\documentclass[aps,prl,reprint,groupedaddress]{revtex4-2}
\usepackage{hyperref}
\usepackage{aas_macros}
\usepackage{graphicx}
\hypersetup{linkcolor=red,citecolor=green,filecolor=cyan,urlcolor=magenta}
\usepackage{xcolor}
\usepackage{amsmath}
\usepackage[resetlabels]{multibib}
\newcites{sec}{Reference}

\begin{document}
	
	
	\title{Gravitational-wave Signature of a First-order Quantum Chromodynamics Phase Transition in Core-Collapse Supernovae}
	
	
	\author{Shuai Zha}
	\email[]{shuai.zha@astro.su.se}
	\affiliation{The Oskar Klein Centre, Department of Astronomy, Stockholm University, AlbaNova, SE-106 91 Stockholm, Sweden}
	
	\author{Evan P. O'Connor}
	\affiliation{The Oskar Klein Centre, Department of Astronomy, Stockholm University, AlbaNova, SE-106 91 Stockholm, Sweden}
	
	\author{Ming-chung Chu}
	\affiliation{Department of Physics and Institute of Theoretical Physics, The Chinese University of Hong Kong, \\Shatin, N.T., Hong Kong S.A.R., China}
	
	\author{Lap-Ming Lin}
	\affiliation{Department of Physics and Institute of Theoretical Physics, The Chinese University of Hong Kong, \\Shatin, N.T., Hong Kong S.A.R., China}
	
	\author{Sean M. Couch}
	\affiliation{Department of Physics and Astronomy, Michigan State University, East Lansing, MI 48824, USA}
	\affiliation{Department of Computational Mathematics, Science, and Engineering, Michigan State University, East Lansing, MI 48824, USA}
	\affiliation{Joint Institute for Nuclear Astrophysics-Center for the Evolution of the Elements, Michigan State University, East Lansing, MI 48824, USA}
	\affiliation{National Superconducting Cyclotron Laboratory, Michigan State University, East Lansing, MI 48824, USA}
	
	\date{\today}
	
	\begin{abstract}
	A first-order quantum chromodynamics (QCD) phase transition (PT) may take place in the protocompact star (PCS)
	produced by a core-collapse supernova (CCSN). In this work, we study the consequences of such a PT in a non-rotating CCSN
	with axisymmetric hydrodynamic simulations. We find that the PT leads to the collapse of the PCS and results in a loud burst of gravitational waves (GWs). The amplitude of this GW burst is $\sim30$ times larger than the post-bounce GW signal normally found for non-rotating CCSN. It shows a broad peak at high frequencies ($\sim2500-4000$ Hz) in the spectrum, has a duration of $\lesssim5 {\rm ms}$, and carries $\sim3$ orders of magnitude more energy than the other episodes. Also, the peak frequency of the PCS oscillation increases dramatically after the PT-induced collapse. In addition to a second neutrino burst, the GW signal, if detected by the ground-based GW detectors, is decisive evidence of the first-order QCD PT inside CCSNe and provides key information about the structure and dynamics of the PCS.
\end{abstract}


\maketitle

\section{Introduction \label{sec:intro}}
Quarks are confined in hadrons such as protons and neutrons at low temperatures and densities. Nonetheless, free quarks should exist in the early universe when the temperature is extremely high ($k_\mathrm{B} T\gtrsim150$ MeV) \cite{1984PhRvD..30..272W}. They may also exist in the cold and superdense interior of compact stars with a density above the nuclear saturation density ($\rho_{\rm sat}\simeq2.6\times10^{14}$ g cm$^{-3}$) \cite[see e.g.,][]{1984PhRvD..30.2379F,1986A&A...160..121H,2013PhRvD..88h3013A}. Moreover, a first-order quantum chromodynamics (QCD) phase transition (PT), i.e., hadron-quark PT, may take place in the protocompact stars (PCS) produced by a core-collapse supernova (CCSN) \cite{2009PhRvL.102h1101S,2013A&A...558A..50N,2018NatAs...2..980F} or binary neutron-star merger \cite{2019PhRvL.122f1101M,2019PhRvL.122f1102B}. Such a PT can result in a more compact PCS and even collapse of the PCS to a black hole (BH). For CCSNe, this can provide an additional energy source for the explosion \cite{2018NatAs...2..980F}, and leads to interesting observational consequences, such as a second neutrino burst \cite{2009PhRvL.102h1101S} and the production of rare $r-$process elements \cite{2020ApJ...894....9F}. 

A galactic CCSN is a yet-undiscovered candidate gravitational-wave (GW) source for ground-based GW detectors \cite{2016PhRvD..94j2001A}. The GWs from a CCSN, in combination with the neutrino and electromagnetic signals, will boost our understanding of the CCSN explosion mechanism \cite{2009CQGra..26t4015O}. Sophisticated multi-dimensional simulations have predicted the GW signals emitted by rotational CCSNe, the oscillations of the proto-neutron stars (PNS), and the standing accretion shock instability \cite[see, e.g., ][]{2007PhRvL..98y1101D,2011LRR....14....1F,2016PhRvL.116o1102H,2018ApJ...865...81O,2019ApJ...876L...9R}. In the meantime, the collapse of a neutron star (NS) to a quark star has been studied in \cite{2006ApJ...639..382L,2009MNRAS.392...52A,2009MNRAS.396.2269D} and GW emission is also found in this scenario. However, in these studies the collapse is triggered artificially by using different equations of state (EOS) for the construction of NS and the hydrodynamic simulation. It is unclear whether the PT-induced collapse of PCS in CCSNe can leave an imprint in the GW signal. In this Letter, we demonstrate the effects of a first-order QCD PT on the GW signal from a non-rotating CCSN, with two-dimensional simulations and a simplified hybrid EOS including hadrons and quarks.

\section{Methods}
\subsection{Equation of state}
To study a first-order QCD PT in CCSNe, we use a hybrid EOS from Ref.~\cite{2009PhRvL.102h1101S,2010JPhG...37i4064S,compose}. This EOS employs the STOS EOS \cite{1998PThPh.100.1013S} for
the hadronic phase and the MIT bag model EOS \cite{1984PhRvD..30.2379F} for the quark phase,
with the Gibbs construction for the mixed phase \cite{1992PhRvD..46.1274G}. The Bag constant
has a value of $B=(165~{\rm MeV})^4$. The Gibbs construction enforces global charge neutrality; therefore, it allows different charge fractions for the hadronic and quark portions in the mixed phase. This leads to a smooth transition from the hadronic phase to the mixed phase (at around $\rho_{\rm sat}$ for the composition of a CCSN core), and the pressure is continuously increasing. A pure quark phase is realized at densities higher than $\sim 3.5\rho_{\rm sat}$, which is similar to the values of those more sophisticated hybrid EOSs \cite{2018NatAs...2..980F,2019PhRvL.122f1102B}. More information about the hybrid EOS can be found in Ref.~\cite{2009PhRvL.102h1101S,1998PThPh.100.1013S,2010JPhG...37i4064S}. 

For this hybrid EOS, there exists a stable branch of the hot third family of compact stars \cite{2013PhRvD..88h3013A}  with a pure quark core (see Fig. 2 in \cite{2016PhRvD..94j3001H}) for matter properties similar to the CCSN core (entropy $\simeq3~{k_{\rm B}/{\rm baryon}}$ and lepton number fraction $\sim0.4$). The maximum mass for the third family ($\sim1.50~M_\odot$) is larger than that of the second family whose core is in the mixed phase. As we will see, this unique property is important for the dynamics of the PCS.

\subsection{FLASH simulation}
We carry out CCSN simulations in two dimensions with the assumption of axisymmetry, using the FLASH code \cite{2000ApJS..131..273F} with an ``M1" scheme for the neutrino transport
\cite{2018ApJ...854...63O}. We take the 12-$M_\odot$, solar-metallicity, presupernova
progenitor $s12$ from \cite{2002RvMP...74.1015W} as the initial conditions. To apply general-relativistic approximations, gravity is calculated with the Case A formulation of \cite{2006A&A...445..273M}. Unlike previous simulations with the FLASH code, we include the lapse function in the Euler equations to mimic the time-dilation effect in general relativity (see the modified Euler equations and some numerical tests in \cite{SI}, also see \cite{2020MNRAS.492.4613O}). This is found to affect the GW frequency significantly after the PT-induced collapse. A cylindrical grid with adaptive mesh refinement is used. It extends out to $2\times10^9$ cm in radius and $\pm2\times10^9$ cm along the cylindrical axis, with a finest resolution of 150 m. We extract the plus GW strain $h_+$ from our Newtonian simulation using the standard quadrupole formula \cite{1990ApJ...351..588F}.

\section{Results}

\subsection{Dynamics}
\begin{figure}[t!]
	\includegraphics[width=0.45\textwidth]{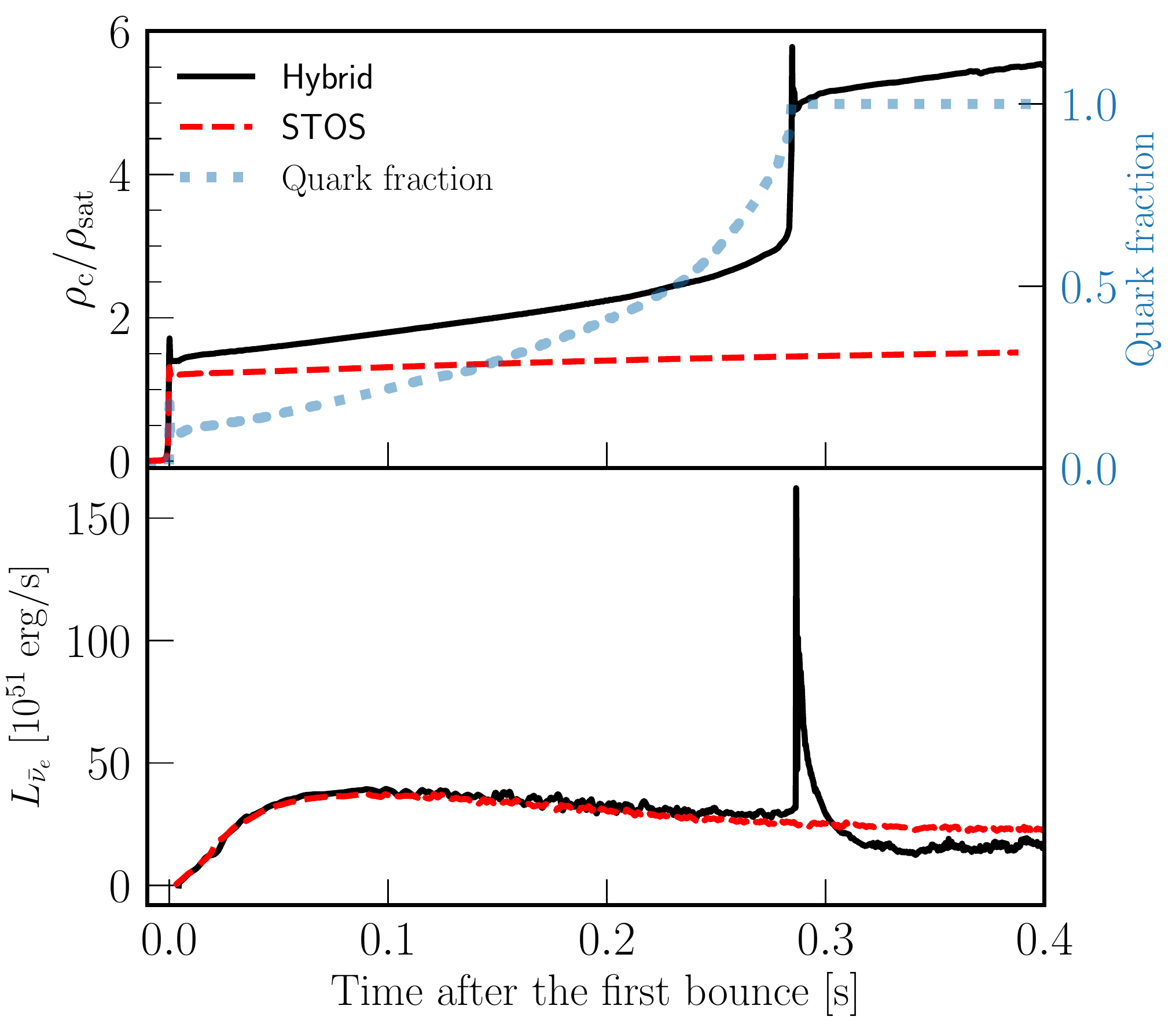}
	\caption{Upper panel: Evolution of central density $\rho_{\rm c}$ after the first bounce for simulations using the hybrid (black solid) and STOS (red dashed) EOSs. Also shown is the quark fraction at the center of the PCS with the hybrid EOS (blue dashed). Lower panel: Luminosity curves of the electron anti-neutrino ($\bar{\nu}_e$) for the two simulations. \label{fig:dyn}}
\end{figure}

To show the consequences of the PT, we run two simulations with the same settings, one using the hybrid EOS and the other using the STOS EOS. The resulting dynamics are shown in the upper panel of Fig.~\ref{fig:dyn}. The iron core of the $s12$ model collapses to above $\rho_{\rm sat}$ and bounces at $t_{\rm b}\simeq151$ ms for both EOSs. At $t_b$, the core of the hybrid EOS has already entered the mixed phase with a central quark mass fraction $X_q=18\%$. However, because the hybrid EOS transitions smoothly from the pure hadronic phase to the mixed phase, the PCS remains in the mixed phase with a low $X_q$ shortly after $t_b$. The bounce shock turns into an accretion shock and stalls at $\sim150$ km at $t_{\rm b}+50$ ms and begins receding inward. 

During the accretion phase, the central density $\rho_{\rm c}$ of the PCS with the hybrid EOS is always larger than that of the PNS with the STOS EOS and $X_q$ continuously increases. The mass of the PCS grows and reaches the maximum of the second family for the hybrid EOS at $\sim t_{\rm b}+286$ ms. The PCS becomes unstable against gravity and experiences a second dynamical collapse. The central density $\rho_{\rm c}$ grows to $1.5\times10^{15}~\rm g~cm^{-3}$ ($\sim6\rho_{\rm sat}$) and the PCS core enters the pure quark phase ($X_q=1$). 

The pure quark core bounces in less than 1 ms at $t_{\rm 2b}$ as the PCS enters the new stable branch of the third family and $\rho_{\rm c}$ drops to $\sim5\rho_{\rm sat}$.
This bounce shock expands quickly to explode the outer envelope. At the end of the simulation, the mean shock radius extends out to $\sim$1500 km with an explosion energy of $\sim2.0\times10^{50}$~erg. The PT-induced collapse is associated with a second neutrino burst with more electron anti-neutrinos than electron neutrinos (lower panel of Fig.~\ref{fig:dyn}), which is consistent with the results of the spherically-symmetric simulation in \cite{2009PhRvL.102h1101S}.

\subsection{Gravitational waves}
\begin{figure}[t!]
	\includegraphics[width=0.48\textwidth]{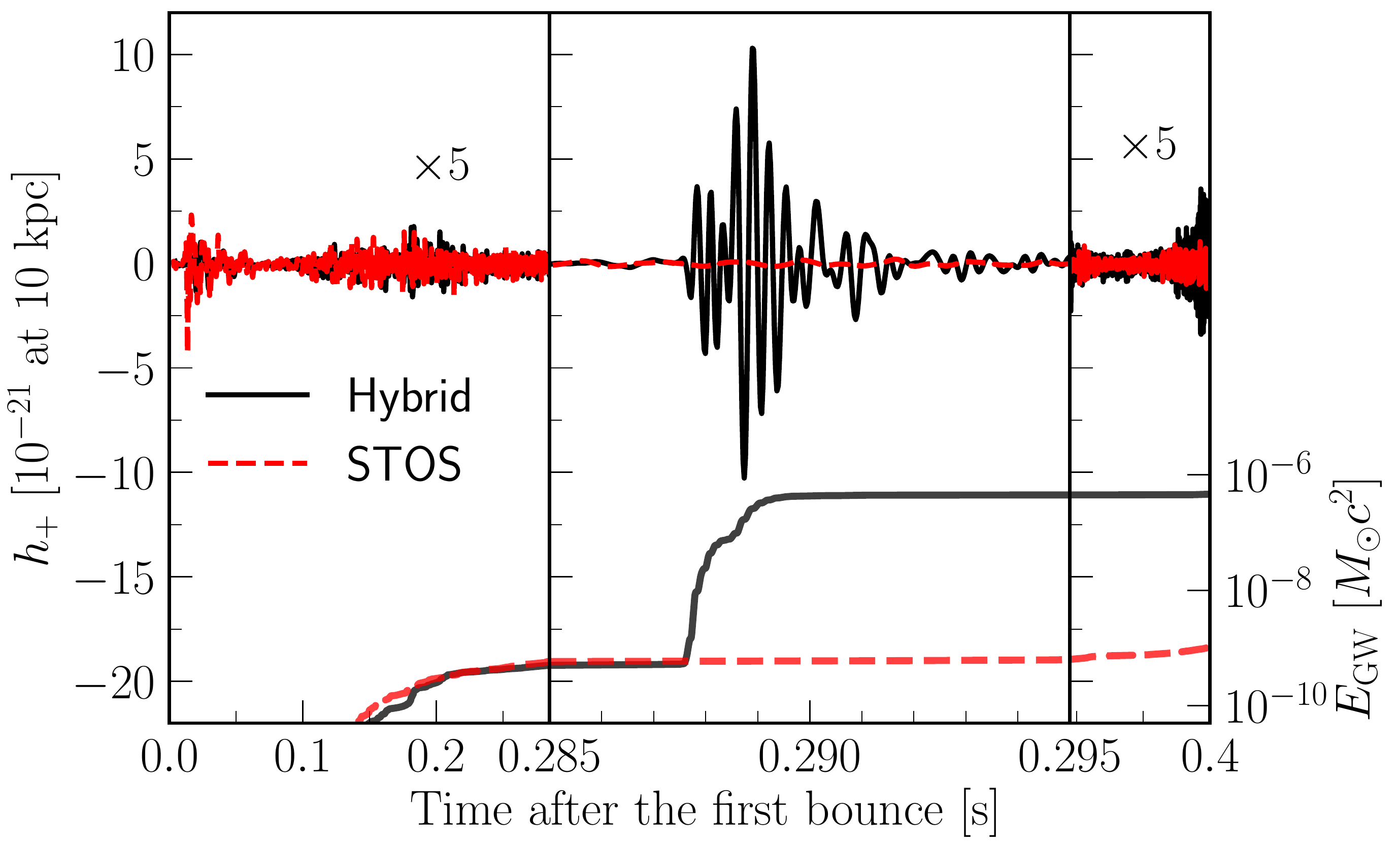}
	\caption{GW waveforms $h_+(t)$ extracted from the CCSN simulations using the hybrid (black solid) and STOS (red dashed) EOSs. The 10 ms window around the second bounce $t_{\rm 2b}\simeq t_{\rm b}+287$ ms is stretched in time to clearly show the loud GW burst. $h_+$ in the other two episodes are multiplied by a factor of 5 to emphasize the contrast in the amplitude. The cumulative emitted GW energies are shown at the bottom.  \label{fig:gw1}}
\end{figure}

In Fig.~\ref{fig:gw1} we show the GW waveforms $h_+(t)$ up to 400 ms after the first bounce ($\sim t_{\rm 2b}+113$ ms) extracted from both simulations. We assume that the distance from the source is 10~kpc. The signal from $t_{\rm b}$ to $\sim t_{\rm b}+50$ ms comes from the prompt convection behind the stalling accretion shock. It is followed by an episode of continuous emission from the oscillations of the PCSs \cite{2019PhRvL.123e1102T}. There is no qualitative difference between the two waveforms until $t_{\rm 2b}$ and the cumulative emitted GW energies are quantitatively similar (bottom of Fig.~\ref{fig:gw1}). In accord with the more compact PCS, the peak GW frequency for the hybrid EOS is always higher than that for the STOS EOS.

Around $t_{\rm 2b}$, the PT-induced collapse results in a burst of GW emission with a much larger amplitude than those of earlier episodes. In Fig.~\ref{fig:gw1}, the 10 ms window around $t_{\rm 2b}$ is stretched in time to show clearly this burst, which is associated with the PT-induced collapse and bounce. The maximum amplitude of $h_+$ reaches $10^{-20}$ and $\sim30$ times larger than those of the other episodes. The energy carried by this burst is $\sim4.6\times10^{-7}~M_\odot c^2$, which is $\sim3$ orders of magnitudes more than the GW energy of the other episodes (and also that of the signal for the STOS EOS). Our numerical test shows that this GW burst results from asphericities developed between $t_{\rm b}$ and $t_{\rm 2b}$ \cite{SI}. After this burst, the amplitude damps quickly to the same level as before $t_{\rm 2b}$. This part of signal should come from the oscillations of the PCS with a pure quark core. 

\begin{figure}[t!]
	\includegraphics[width=0.48\textwidth]{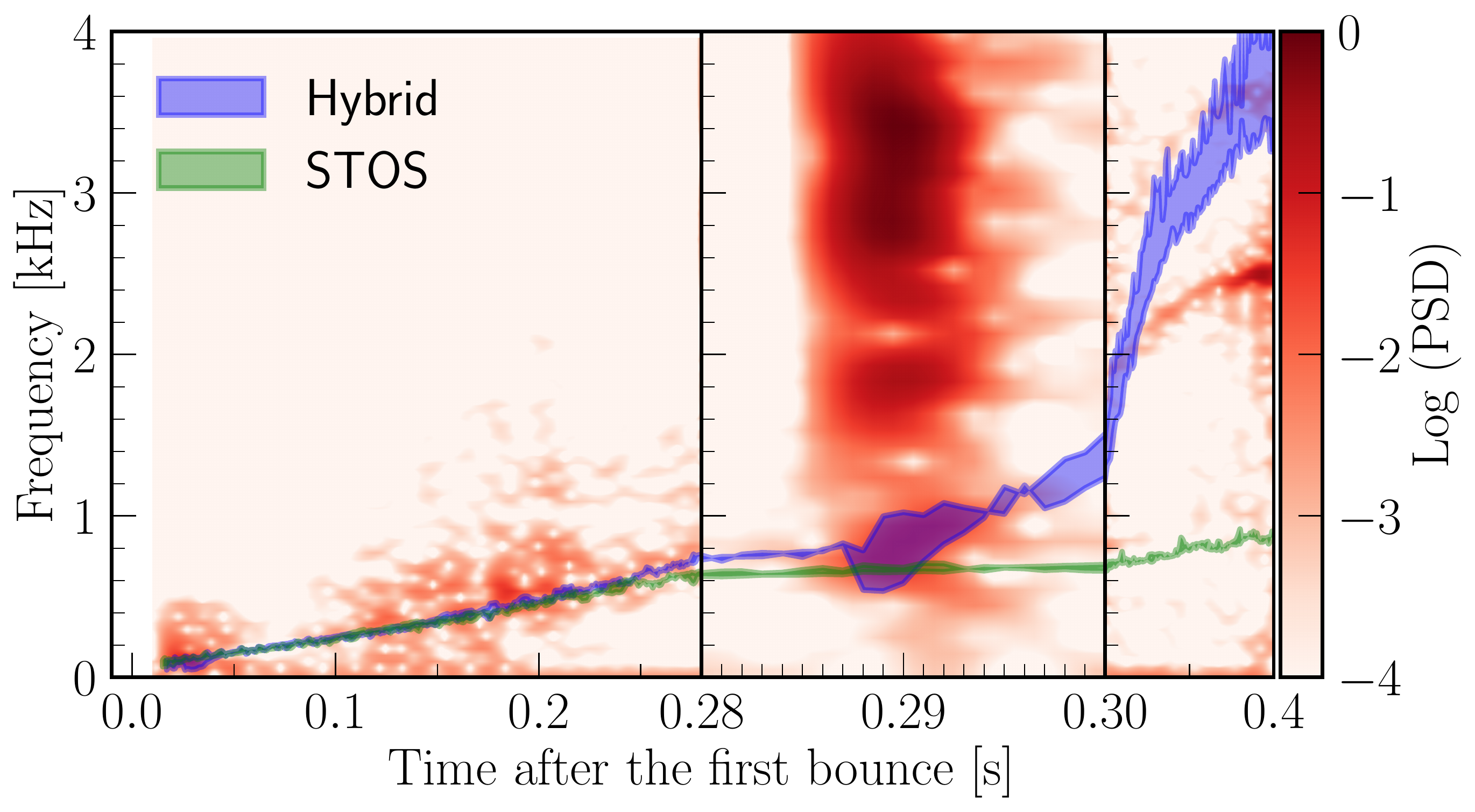}
	\caption{Colormap of the time-dependent power spectral density (PSD) for the GW waveform extracted from the simulation using the hybrid EOS. The color-filled bands track the evolution of the Brunt-V\"ais\"al\"a frequencies at densities between $10^{11}$ and $10^{12}$ g cm$^{-3}$ for the hybrid (blue) and STOS (green) EOSs. \label{fig:gw2}}
\end{figure}

A time-dependent spectrum (or spectrogram) is useful for understanding the emission mechanisms of GWs, as well as designing efficient detection strategies. Fig.~\ref{fig:gw2} shows the spectrogram of the GW signal extracted from the simulation using the hybrid EOS. We use a Kaiser window with a width of 25~ms for the short-time Fourier transform except for around $t_{\rm 2b}$ where a width of 10~ms is used. Before $t_{\rm 2b}$, the spectral evolution is similar to that of the STOS EOS (see \cite{SI}). The GW peak frequency is continuously increasing, in accord with the evolution of the Brunt-V\"ais\"al\"a frequencies $f_{\rm BV}$ (Eq. (3) in \cite{SI}) at densities between $10^{11}$ and $10^{12}$ g cm$^{-3}$ (blue band in Fig.~\ref{fig:gw2}), which is approximately the PCS surface \cite{2013ApJ...766...43M}.  Around $t_{\rm 2b}$, the GW burst has a much higher frequency ($\sim2500-4000$~Hz). This is related to the change of the dominant GW emission region from $\sim10-20$ km to $\sim5-10$ km (see \cite{SI}), which is inside the quasi-static core during the second collapse and bounce. During this time, $f_{\rm BV}$ peaks at 2900 Hz near a radius of $10$ km ($\rho\simeq8\times10^{13}$ g~cm$^{-3}$) and is closer to the observed GW frequency. 

Shortly after $t_{\rm 2b}$, the GW peak frequency drops back to $\sim1000$ Hz and continues to increase afterwards, albeit at a much faster rate. We find that $f_{\rm BV}$ has a much larger spread inside the PCS and the GW spectral evolution does not match the track of $f_{\rm BV}$. After $t_{\rm b}+300$ ms, the peak frequency of the dominant GW emission is closer to $f_{\rm BV}$ at densities $\sim5\times10^{12}$ g~cm$^{-3}$. Nevertheless, due to the much larger $\rho_{\rm c}$ ($\gtrsim4$ times) and compactness of the PCS for the hybrid EOS (Fig.~\ref{fig:dyn}), the peak GW frequency is $2-3$ times higher than that for the STOS EOS. The GW spectral evolution after $t_{\rm 2b}$ contains information about the structure and evolution of the PCS with a pure quark core, from which one may infer the properties of the quark EOS (e.g. Bag constant).

\subsection{Detection prospect}
To estimate the detectability of the GW signals, we calculate the dimensionless characteristic GW strain ($h_{\rm char}$) \cite{1998PhRvD..57.4535F} assuming a distance of 10 kpc, and compare it with the sensitivity of Advanced LIGO in Fig.~\ref{fig:hchar}. Below $\sim 1000$ Hz, $h_{\rm char}$ are quantitatively similar for the hybrid and STOS EOSs. At higher frequencies, $h_{\rm char}$ for the hybrid EOS shows a broad peak between $\sim2500-4000~{\rm Hz}$, which is also above the detector's sensitivity curve. This part is mainly contributed by the burst associated with the PT-induced collapse, seen from the comparison between the entire $h_{\rm char}$ and that between $t_{\rm 2b}-3 {\rm ms}$ and $t_{\rm 2b}+7 {\rm ms}$.

We calculate the single-detector signal-to-noise ratio (SNR)  of the GW waveforms assuming the optimal orientation using Eq.~(1.1) in \cite{1998PhRvD..57.4535F}. If a confident detection requires an SNR of 8, then for the hybrid EOS, inclusion (exclusion) of the burst yields a detection radius of 22 (12) kpc. The detectability of the burst is not significantly better because the current detectors are optimized for GW signals at $\sim10-1000$ Hz. The amplitude of $h_{\rm char}$ and detector noise increase together by a factor of 10 from 100 Hz to 3000 Hz (Fig.~\ref{fig:hchar}). Future-generation detectors, such as the Einstein Telescope \cite{2010CQGra..27s4002P} and the Cosmic Explorer \cite{2017CQGra..34d4001A}, may consider improving the sensitivity at several kHz if such signals are targeted (also for BH forming CCSNe \cite{2018ApJ...857...13P}). 

\begin{figure}[t!]
	\includegraphics[width=0.48\textwidth]{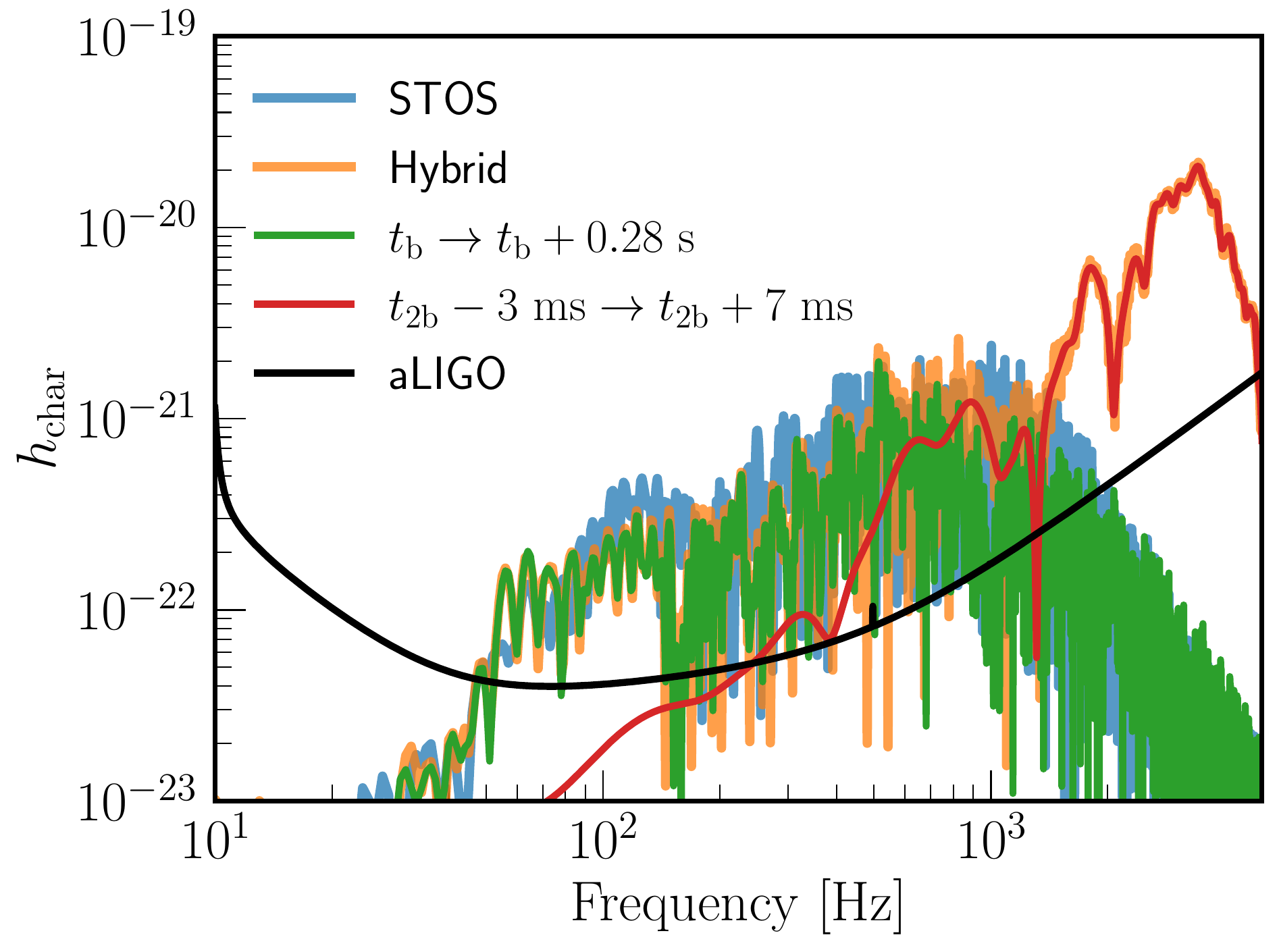}
	\caption{Dimension-less characteristic strain spectra ($h_{\rm char}(f)$) of the GW signals from $t_{\rm b}$ to $t_{\rm b}+400$ ms with the STOS (blue) and hybrid (orange) EOSs. Also shown are the $h_{\rm char}(f)$ of the GW signal for the hybrid EOS in the time interval of: from $t_{\rm b}$ to $t_{\rm b}+280$ ms (green) and between $t_{\rm 2b}-3 {\rm ms}$ and $t_{\rm 2b}+7 {\rm ms} $ ms (red). The black line is the sensitivity spectrum of Advanced LIGO \cite{LIGO_v5}. \label{fig:hchar}}
\end{figure}

In this work, the GW waveforms $h_+(t)$ are extracted from 2D simulations with the assumption of axisymmetry. In various studies \cite[e.g., ][]{2018ApJ...865...81O,2017MNRAS.468.2032A,2019ApJ...876L...9R}, the amplitude of $h_+$ in 3D simulations can be $5-10$ times smaller than those in their 2D counterparts, which lowers the expectation for the GW detection. Future study is needed to explore 3D effects for the PT-induced collapse and observables. We expect that the burst associated with the PT-induced collapse would still be present in 3D simulations but with smaller amplitudes.

\section{Discussion}

We present here a specific case in which the PT-induced collapse results in a bounce shock that successfully explodes the mantle. However, for other progenitors \cite{2011ApJ...730...70O} or other hybrid EOSs \cite{2009PhRvL.102h1101S}, the star may fail to explode and collapse into a black hole (BH) in two scenarios. First, the PCS at the onset of the PT-induced collapse may exceed the maximum mass which the hybrid EOS permits and it directly collapses into a BH. In this case, the GW (and neutrino) burst reported here will be absent. Nevertheless, the existence of free quarks in the PCS might be inferred from the shortening of the BH formation time \cite{2013A&A...558A..50N,2020ApJ...894....4S}, though it is subjected to the uncertainties of the pure hadronic EOS.

In the second scenario, the second bounce shock is launched but the PCS still collapses into a BH at a later time. In this case, the burst of GWs and neutrinos associated with the PT-induced collapse will be present, followed by the shut-off of both signals at BH formation. This is an interesting possibility to be explored. Moreover, in both cases of BH formation, if the iron core is rapidly rotating, the inclusion of PT can produce different BH ring-down signals compared to those for a hadronic EOS due to the different free-fall time of the PCS, which is found in binary NS-merger simulations \cite{2019PhRvL.122f1101M}.


\section{Conclusions}
In this Letter, we demonstrate the effects of a first-order QCD PT on the GW signals from a non-rotating CCSN. We find that the PT results in the collapse of the PCS  at $\rho_{\rm c}\sim3.5\rho_{\rm sat}$, and the core radiates a loud GW burst in $\lesssim5$ ms. The amplitude of this burst reaches $h_+=10^{-20}$ assuming a source distance of 10 kpc and is larger by a factor of $\sim30$ than other episodes of GW emission (and generally those using a hadronic EOS). The spectrum of this burst shows a broad peak at $\sim2500-4000$ Hz, which is higher than that generally found for CCSNe without the PT-induced collapse. The peak GW frequency following this burst is also much higher ($>$1 kHz) than that for the hadronic EOS due to the large compactness of the PCS with a pure quark core. Therefore, the PT inside a CCSN can be inferred from the GW detection. However, the louder burst is not necessarily easier to detect because of the increasing noise level of high frequencies for current ground based GW detectors. Nevertheless, the loud, high-frequency burst of GW radiation over a short period of time may be a prime target for future searches of coherent wave burst signals \cite{2008CQGra..25k4029K}.

The hybrid EOS transitions from the hadronic phase to the mixed phase at a low density ($\sim\rho_{\rm sat}$). Ref.~\cite{2018NatAs...2..980F} simulated CCSNe in spherical symmetry using a more physical and complex hybrid EOS (DD2F-SF, transition density $\sim2.4\rho_{\rm sat}$), and the dynamics of the second collapse are similar to our results. Therefore, we expect that the properties of the GW burst (i.e. the frequency and amplitude) associated with the PT-induced collapse should still be present with a more physical EOS, such as DD2F-SF \cite{2018NatAs...2..980F,2019PhRvL.122f1102B} or a Chiral Mean Field model \cite{2019PhRvL.122f1101M}, which are consistent with the maximum NS mass measurement \cite{2013Sci...340..448A}. A natural extension is to employ such EOSs in multi-dimensional CCSN simulations. Moreover, progenitor dependence, such as the initial mass and rotation, should be studied to acquire a more comprehensive picture for the effects of a first-order QCD PT on the GWs from CCSNe. Particularly, we expect that if the iron core is rapidly rotating before collapse, the GW burst associated with the PT-induced collapse will be much louder and it may allow the detection for sources farther away.

\begin{acknowledgments}
	We thank I. Sagert for making the hybrid equation of state publicly available at the CompOSE website \cite{compose}. This work is supported by the Swedish  Research  Council (Project No. 2018-04575) and by the Research Grant
	Council of Hong Kong (Project 14300317). The simulations were performed on resources provided by the Swedish  National  Infrastructure  for  Computing  (SNIC)  at PDC and NSC. We also acknowledge the support of the CUHK Central High Performance Computing
	Cluster, on which parts of the simulations were performed. 
\end{acknowledgments}

\bibliography{the_bib}

\begin{thebibliography}{46}%
\makeatletter
\providecommand \@ifxundefined [1]{%
 \@ifx{#1\undefined}
}%
\providecommand \@ifnum [1]{%
 \ifnum #1\expandafter \@firstoftwo
 \else \expandafter \@secondoftwo
 \fi
}%
\providecommand \@ifx [1]{%
 \ifx #1\expandafter \@firstoftwo
 \else \expandafter \@secondoftwo
 \fi
}%
\providecommand \natexlab [1]{#1}%
\providecommand \enquote  [1]{``#1''}%
\providecommand \bibnamefont  [1]{#1}%
\providecommand \bibfnamefont [1]{#1}%
\providecommand \citenamefont [1]{#1}%
\providecommand \href@noop [0]{\@secondoftwo}%
\providecommand \href [0]{\begingroup \@sanitize@url \@href}%
\providecommand \@href[1]{\@@startlink{#1}\@@href}%
\providecommand \@@href[1]{\endgroup#1\@@endlink}%
\providecommand \@sanitize@url [0]{\catcode `\\12\catcode `\$12\catcode
  `\&12\catcode `\#12\catcode `\^12\catcode `\_12\catcode `\%12\relax}%
\providecommand \@@startlink[1]{}%
\providecommand \@@endlink[0]{}%
\providecommand \url  [0]{\begingroup\@sanitize@url \@url }%
\providecommand \@url [1]{\endgroup\@href {#1}{\urlprefix }}%
\providecommand \urlprefix  [0]{URL }%
\providecommand \Eprint [0]{\href }%
\providecommand \doibase [0]{https://doi.org/}%
\providecommand \selectlanguage [0]{\@gobble}%
\providecommand \bibinfo  [0]{\@secondoftwo}%
\providecommand \bibfield  [0]{\@secondoftwo}%
\providecommand \translation [1]{[#1]}%
\providecommand \BibitemOpen [0]{}%
\providecommand \bibitemStop [0]{}%
\providecommand \bibitemNoStop [0]{.\EOS\space}%
\providecommand \EOS [0]{\spacefactor3000\relax}%
\providecommand \BibitemShut  [1]{\csname bibitem#1\endcsname}%
\let\auto@bib@innerbib\@empty
\bibitem [{\citenamefont {{Witten}}(1984)}]{1984PhRvD..30..272W}%
  \BibitemOpen
  \bibfield  {author} {\bibinfo {author} {E.~{Witten}},\ }\href
  {https://doi.org/10.1103/PhysRevD.30.272} {\bibfield  {journal} {\bibinfo
  {journal} {\prd}\ }\textbf {\bibinfo {volume} {30}},\ \bibinfo {pages} {272}
  (\bibinfo {year} {1984})}\BibitemShut {NoStop}%
\bibitem [{\citenamefont {{Farhi}}\ and\ \citenamefont
  {{Jaffe}}(1984)}]{1984PhRvD..30.2379F}%
  \BibitemOpen
  \bibfield  {author} {\bibinfo {author} {E.~{Farhi}}\ and\ \bibinfo {author}
  {R.~L. {Jaffe}},\ }\href {https://doi.org/10.1103/PhysRevD.30.2379}
  {\bibfield  {journal} {\bibinfo  {journal} {\prd}\ }\textbf {\bibinfo
  {volume} {30}},\ \bibinfo {pages} {2379} (\bibinfo {year}
  {1984})}\BibitemShut {NoStop}%
\bibitem [{\citenamefont {{Haensel}}\ \emph {et~al.}(1986)\citenamefont
  {{Haensel}}, \citenamefont {{Zdunik}},\ and\ \citenamefont
  {{Schaefer}}}]{1986A&A...160..121H}%
  \BibitemOpen
  \bibfield  {author} {\bibinfo {author} {P.~{Haensel}}, \bibinfo {author}
  {J.~L. {Zdunik}},\ and\ \bibinfo {author} {R.~{Schaefer}},\ }\href@noop {}
  {\bibfield  {journal} {\bibinfo  {journal} {\aap}\ }\textbf {\bibinfo
  {volume} {160}},\ \bibinfo {pages} {121} (\bibinfo {year}
  {1986})}\BibitemShut {NoStop}%
\bibitem [{\citenamefont {{Alford}}\ \emph {et~al.}(2013)\citenamefont
  {{Alford}}, \citenamefont {{Han}},\ and\ \citenamefont
  {{Prakash}}}]{2013PhRvD..88h3013A}%
  \BibitemOpen
  \bibfield  {author} {\bibinfo {author} {M.~G. {Alford}}, \bibinfo {author}
  {S.~{Han}},\ and\ \bibinfo {author} {M.~{Prakash}},\ }\href
  {https://doi.org/10.1103/PhysRevD.88.083013} {\bibfield  {journal} {\bibinfo
  {journal} {\prd}\ }\textbf {\bibinfo {volume} {88}},\ \bibinfo {eid} {083013}
  (\bibinfo {year} {2013})},\ \Eprint {https://arxiv.org/abs/1302.4732}
  {arXiv:1302.4732 [astro-ph.SR]} \BibitemShut {NoStop}%
\bibitem [{\citenamefont {{Sagert}}\ \emph {et~al.}(2009)\citenamefont
  {{Sagert}}, \citenamefont {{Fischer}}, \citenamefont {{Hempel}},
  \citenamefont {{Pagliara}}, \citenamefont {{Schaffner-Bielich}},
  \citenamefont {{Mezzacappa}}, \citenamefont {{Thielemann}},\ and\
  \citenamefont {{Liebend{\"o}rfer}}}]{2009PhRvL.102h1101S}%
  \BibitemOpen
  \bibfield  {author} {\bibinfo {author} {I.~{Sagert}}, \bibinfo {author}
  {T.~{Fischer}}, \bibinfo {author} {M.~{Hempel}}, \bibinfo {author}
  {G.~{Pagliara}}, \bibinfo {author} {J.~{Schaffner-Bielich}}, \bibinfo
  {author} {A.~{Mezzacappa}}, \bibinfo {author} {F.~K. {Thielemann}},\ and\
  \bibinfo {author} {M.~{Liebend{\"o}rfer}},\ }\href
  {https://doi.org/10.1103/PhysRevLett.102.081101} {\bibfield  {journal}
  {\bibinfo  {journal} {\prl}\ }\textbf {\bibinfo {volume} {102}},\ \bibinfo
  {eid} {081101} (\bibinfo {year} {2009})},\ \Eprint
  {https://arxiv.org/abs/0809.4225} {arXiv:0809.4225 [astro-ph]} \BibitemShut
  {NoStop}%
\bibitem [{\citenamefont {{Nakazato}}\ \emph {et~al.}(2013)\citenamefont
  {{Nakazato}}, \citenamefont {{Sumiyoshi}},\ and\ \citenamefont
  {{Yamada}}}]{2013A&A...558A..50N}%
  \BibitemOpen
  \bibfield  {author} {\bibinfo {author} {K.~{Nakazato}}, \bibinfo {author}
  {K.~{Sumiyoshi}},\ and\ \bibinfo {author} {S.~{Yamada}},\ }\href
  {https://doi.org/10.1051/0004-6361/201322231} {\bibfield  {journal} {\bibinfo
   {journal} {\aap}\ }\textbf {\bibinfo {volume} {558}},\ \bibinfo {eid} {A50}
  (\bibinfo {year} {2013})},\ \Eprint {https://arxiv.org/abs/1309.3383}
  {arXiv:1309.3383 [astro-ph.HE]} \BibitemShut {NoStop}%
\bibitem [{\citenamefont {{Fischer}}\ \emph {et~al.}(2018)\citenamefont
  {{Fischer}}, \citenamefont {{Bastian}}, \citenamefont {{Wu}}, \citenamefont
  {{Baklanov}}, \citenamefont {{Sorokina}}, \citenamefont {{Blinnikov}},
  \citenamefont {{Typel}}, \citenamefont {{Kl{\"a}hn}},\ and\ \citenamefont
  {{Blaschke}}}]{2018NatAs...2..980F}%
  \BibitemOpen
  \bibfield  {author} {\bibinfo {author} {T.~{Fischer}}, \bibinfo {author}
  {N.-U.~F. {Bastian}}, \bibinfo {author} {M.-R. {Wu}}, \bibinfo {author}
  {P.~{Baklanov}}, \bibinfo {author} {E.~{Sorokina}}, \bibinfo {author}
  {S.~{Blinnikov}}, \bibinfo {author} {S.~{Typel}}, \bibinfo {author}
  {T.~{Kl{\"a}hn}},\ and\ \bibinfo {author} {D.~B. {Blaschke}},\ }\href
  {https://doi.org/10.1038/s41550-018-0583-0} {\bibfield  {journal} {\bibinfo
  {journal} {Nat. Astron.}\ }\textbf {\bibinfo {volume} {2}},\ \bibinfo {pages}
  {980} (\bibinfo {year} {2018})},\ \Eprint {https://arxiv.org/abs/1712.08788}
  {arXiv:1712.08788 [astro-ph.HE]} \BibitemShut {NoStop}%
\bibitem [{\citenamefont {{Most}}\ \emph {et~al.}(2019)\citenamefont {{Most}},
  \citenamefont {{Papenfort}}, \citenamefont {{Dexheimer}}, \citenamefont
  {{Hanauske}}, \citenamefont {{Schramm}}, \citenamefont {{St{\"o}cker}},\ and\
  \citenamefont {{Rezzolla}}}]{2019PhRvL.122f1101M}%
  \BibitemOpen
  \bibfield  {author} {\bibinfo {author} {E.~R. {Most}}, \bibinfo {author}
  {L.~J. {Papenfort}}, \bibinfo {author} {V.~{Dexheimer}}, \bibinfo {author}
  {M.~{Hanauske}}, \bibinfo {author} {S.~{Schramm}}, \bibinfo {author}
  {H.~{St{\"o}cker}},\ and\ \bibinfo {author} {L.~{Rezzolla}},\ }\href
  {https://doi.org/10.1103/PhysRevLett.122.061101} {\bibfield  {journal}
  {\bibinfo  {journal} {\prl}\ }\textbf {\bibinfo {volume} {122}},\ \bibinfo
  {eid} {061101} (\bibinfo {year} {2019})},\ \Eprint
  {https://arxiv.org/abs/1807.03684} {arXiv:1807.03684 [astro-ph.HE]}
  \BibitemShut {NoStop}%
\bibitem [{\citenamefont {{Bauswein}}\ \emph {et~al.}(2019)\citenamefont
  {{Bauswein}}, \citenamefont {{Bastian}}, \citenamefont {{Blaschke}},
  \citenamefont {{Chatziioannou}}, \citenamefont {{Clark}}, \citenamefont
  {{Fischer}},\ and\ \citenamefont {{Oertel}}}]{2019PhRvL.122f1102B}%
  \BibitemOpen
  \bibfield  {author} {\bibinfo {author} {A.~{Bauswein}}, \bibinfo {author}
  {N.-U.~F. {Bastian}}, \bibinfo {author} {D.~B. {Blaschke}}, \bibinfo {author}
  {K.~{Chatziioannou}}, \bibinfo {author} {J.~A. {Clark}}, \bibinfo {author}
  {T.~{Fischer}},\ and\ \bibinfo {author} {M.~{Oertel}},\ }\href
  {https://doi.org/10.1103/PhysRevLett.122.061102} {\bibfield  {journal}
  {\bibinfo  {journal} {\prl}\ }\textbf {\bibinfo {volume} {122}},\ \bibinfo
  {eid} {061102} (\bibinfo {year} {2019})},\ \Eprint
  {https://arxiv.org/abs/1809.01116} {arXiv:1809.01116 [astro-ph.HE]}
  \BibitemShut {NoStop}%
\bibitem [{\citenamefont {{Fischer}}\ \emph {et~al.}(2020)\citenamefont
  {{Fischer}}, \citenamefont {{Wu}}, \citenamefont {{Wehmeyer}}, \citenamefont
  {{Bastian}}, \citenamefont {{Mart{\'\i}nez-Pinedo}},\ and\ \citenamefont
  {{Thielemann}}}]{2020ApJ...894....9F}%
  \BibitemOpen
  \bibfield  {author} {\bibinfo {author} {T.~{Fischer}}, \bibinfo {author}
  {M.-R. {Wu}}, \bibinfo {author} {B.~{Wehmeyer}}, \bibinfo {author} {N.-U.~F.
  {Bastian}}, \bibinfo {author} {G.~{Mart{\'\i}nez-Pinedo}},\ and\ \bibinfo
  {author} {F.-K. {Thielemann}},\ }\href
  {https://doi.org/10.3847/1538-4357/ab86b0} {\bibfield  {journal} {\bibinfo
  {journal} {\apj}\ }\textbf {\bibinfo {volume} {894}},\ \bibinfo {eid} {9}
  (\bibinfo {year} {2020})},\ \Eprint {https://arxiv.org/abs/2003.00972}
  {arXiv:2003.00972 [astro-ph.HE]} \BibitemShut {NoStop}%
\bibitem [{\citenamefont {{Abbott}}\ \emph {et~al.}(2016)\citenamefont
  {{Abbott}}, \citenamefont {{Abbott}}, \citenamefont {{Abbott}}, \citenamefont
  {{Abernathy}}, \citenamefont {{Acernese}}, \citenamefont {{Ackley}},
  \citenamefont {{Adams}}, \citenamefont {{Adams}}, \citenamefont {{Addesso}},
  \citenamefont {{Adhikari}}, \citenamefont {{Adya}}, \citenamefont
  {{Affeldt}}, \citenamefont {{Agathos}}, \citenamefont {{Agatsuma}},
  \citenamefont {{Aggarwal}}, \citenamefont {{Aguiar}}, \citenamefont
  {{Aiello}}, \citenamefont {{Ain}}, \citenamefont {{Ajith}}, \citenamefont
  {{Allen}}, \citenamefont {{Allocca}}, \citenamefont {{Altin}}, \citenamefont
  {{Anderson}}, \citenamefont {{Anderson}}, \citenamefont {{Arai}},
  \citenamefont {{Araya}}, \citenamefont {{Arceneaux}}, \citenamefont
  {{Areeda}}, \citenamefont {{Arnaud}}, \citenamefont {{Arun}}, \citenamefont
  {{Ascenzi}}, \citenamefont {{Ashton}}, \citenamefont {{Ast}}, \citenamefont
  {{Aston}}, \citenamefont {{Astone}}, \citenamefont {{Aufmuth}}, \citenamefont
  {{Aulbert}}, \citenamefont {{Babak}}, \citenamefont {{Bacon}}, \citenamefont
  {{Bader}}, \citenamefont {{Baker}}, \citenamefont {{Baldaccini}},
  \citenamefont {{Ballardin}}, \citenamefont {{Ballmer}}, \citenamefont
  {{Barayoga}}, \citenamefont {{Barclay}}, \citenamefont {{Barish}},
  \citenamefont {{Barker}}, \citenamefont {{Barone}}, \citenamefont {{Barr}},
  \citenamefont {{Barsotti}}, \citenamefont {{Barsuglia}}, \citenamefont
  {{Barta}}, \citenamefont {{Bartlett}}, \citenamefont {{Bartos}},
  \citenamefont {{Bassiri}}, \citenamefont {{Basti}}, \citenamefont {{Batch}},
  \citenamefont {{Baune}}, \citenamefont {{Bavigadda}}, \citenamefont
  {{Bazzan}}, \citenamefont {{Behnke}}, \citenamefont {{Bejger}}, \citenamefont
  {{Bell}}, \citenamefont {{Bell}}, \citenamefont {{Berger}}, \citenamefont
  {{Bergman}}, \citenamefont {{Bergmann}}, \citenamefont {{Berry}},
  \citenamefont {{Bersanetti}}, \citenamefont {{Bertolini}}, \citenamefont
  {{Betzwieser}}, \citenamefont {{Bhagwat}}, \citenamefont {{Bhand are}},
  \citenamefont {{Bilenko}}, \citenamefont {{Billingsley}}, \citenamefont
  {{Birch}}, \citenamefont {{Birney}}, \citenamefont {{Biscans}}, \citenamefont
  {{Bisht}}, \citenamefont {{Bitossi}}, \citenamefont {{Biwer}}, \citenamefont
  {{Bizouard}}, \citenamefont {{Blackburn}}, \citenamefont {{Blair}},
  \citenamefont {{Blair}}, \citenamefont {{Blair}}, \citenamefont {{Bloemen}},
  \citenamefont {{Bock}}, \citenamefont {{Bodiya}}, \citenamefont {{Boer}},
  \citenamefont {{Bogaert}}, \citenamefont {{Bogan}}, \citenamefont {{Bohe}},
  \citenamefont {{Bojtos}}, \citenamefont {{Bond}}, \citenamefont {{Bondu}},
  \citenamefont {{Bonnand}}, \citenamefont {{Boom}}, \citenamefont {{Bork}},
  \citenamefont {{Boschi}}, \citenamefont {{Bose}}, \citenamefont
  {{Bouffanais}}, \citenamefont {{Bozzi}}, \citenamefont {{Bradaschia}},
  \citenamefont {{Brady}}, \citenamefont {{Braginsky}}, \citenamefont
  {{Branchesi}}, \citenamefont {{Brau}}, \citenamefont {{Briant}},
  \citenamefont {{Brillet}}, \citenamefont {{Brinkmann}}, \citenamefont
  {{Brisson}}, \citenamefont {{Brockill}}, \citenamefont {{Brooks}},
  \citenamefont {{Brown}}, \citenamefont {{Brown}}, \citenamefont {{Brown}},
  \citenamefont {{Buchanan}}, \citenamefont {{Buikema}}, \citenamefont
  {{Bulik}}, \citenamefont {{Bulten}}, \citenamefont {{Buonanno}},
  \citenamefont {{Buskulic}}, \citenamefont {{Buy}}, \citenamefont {{Byer}},
  \citenamefont {{Cadonati}}, \citenamefont {{Cagnoli}}, \citenamefont
  {{Cahillane}}, \citenamefont {{Calder{\'o}n Bustillo}}, \citenamefont
  {{Callister}}, \citenamefont {{Calloni}}, \citenamefont {{Camp}},
  \citenamefont {{Cannon}}, \citenamefont {{Cao}}, \citenamefont {{Capano}},
  \citenamefont {{Capocasa}}, \citenamefont {{Carbognani}}, \citenamefont
  {{Caride}}, \citenamefont {{Casanueva Diaz}}, \citenamefont {{Casentini}},
  \citenamefont {{Caudill}}, \citenamefont {{Cavagli{\`a}}}, \citenamefont
  {{Cavalier}}, \citenamefont {{Cavalieri}}, \citenamefont {{Cella}},
  \citenamefont {{Cepeda}}, \citenamefont {{Cerboni Baiardi}}, \citenamefont
  {{Cerretani}}, \citenamefont {{Cesarini}}, \citenamefont {{Chakraborty}},
  \citenamefont {{Chalermsongsak}}, \citenamefont {{Chamberlin}}, \citenamefont
  {{Chan}}, \citenamefont {{Chao}}, \citenamefont {{Charlton}}, \citenamefont
  {{Chassande-Mottin}}, \citenamefont {{Chen}}, \citenamefont {{Chen}},
  \citenamefont {{Cheng}}, \citenamefont {{Chincarini}}, \citenamefont
  {{Chiummo}}, \citenamefont {{Cho}}, \citenamefont {{Cho}}, \citenamefont
  {{Chow}}, \citenamefont {{Christensen}}, \citenamefont {{Chu}}, \citenamefont
  {{Chua}}, \citenamefont {{Chung}}, \citenamefont {{Ciani}}, \citenamefont
  {{Clara}}, \citenamefont {{Clark}}, \citenamefont {{Cleva}}, \citenamefont
  {{Coccia}}, \citenamefont {{Cohadon}}, \citenamefont {{Colla}}, \citenamefont
  {{Collette}}, \citenamefont {{Cominsky}}, \citenamefont {{Constancio}},
  \citenamefont {{Conte}}, \citenamefont {{Conti}}, \citenamefont {{Cook}},
  \citenamefont {{Corbitt}}, \citenamefont {{Cornish}}, \citenamefont
  {{Corpuz}}, \citenamefont {{Corsi}}, \citenamefont {{Cortese}}, \citenamefont
  {{Costa}}, \citenamefont {{Coughlin}}, \citenamefont {{Coughlin}},
  \citenamefont {{Coulon}}, \citenamefont {{Countryman}}, \citenamefont
  {{Couvares}}, \citenamefont {{Coward}}, \citenamefont {{Cowart}},
  \citenamefont {{Coyne}}, \citenamefont {{Coyne}}, \citenamefont {{Craig}},
  \citenamefont {{Creighton}}, \citenamefont {{Cripe}}, \citenamefont
  {{Crowder}}, \citenamefont {{Cumming}}, \citenamefont {{Cunningham}},
  \citenamefont {{Cuoco}}, \citenamefont {{Dal Canton}}, \citenamefont
  {{Danilishin}}, \citenamefont {{D'Antonio}}, \citenamefont {{Danzmann}},
  \citenamefont {{Darman}}, \citenamefont {{Dattilo}}, \citenamefont {{Dave}},
  \citenamefont {{Daveloza}}, \citenamefont {{Davier}}, \citenamefont
  {{Davies}}, \citenamefont {{Daw}}, \citenamefont {{Day}}, \citenamefont
  {{DeBra}}, \citenamefont {{Debreczeni}}, \citenamefont {{Degallaix}},
  \citenamefont {{De Laurentis}}, \citenamefont {{Del{\'e}glise}},
  \citenamefont {{Del Pozzo}}, \citenamefont {{Denker}}, \citenamefont
  {{Dent}}, \citenamefont {{Dergachev}}, \citenamefont {{De Rosa}},
  \citenamefont {{DeRosa}}, \citenamefont {{DeSalvo}}, \citenamefont
  {{Dhurandhar}}, \citenamefont {{D{\'\i}az}}, \citenamefont {{Di Fiore}},
  \citenamefont {{Di Giovanni}}, \citenamefont {{Di Girolamo}}, \citenamefont
  {{Di Lieto}}, \citenamefont {{Di Pace}}, \citenamefont {{Di Palma}},
  \citenamefont {{Di Virgilio}}, \citenamefont {{Dojcinoski}}, \citenamefont
  {{Dolique}}, \citenamefont {{Donovan}}, \citenamefont {{Dooley}},
  \citenamefont {{Doravari}}, \citenamefont {{Douglas}}, \citenamefont
  {{Downes}}, \citenamefont {{Drago}}, \citenamefont {{Drever}}, \citenamefont
  {{Driggers}}, \citenamefont {{Du}}, \citenamefont {{Ducrot}}, \citenamefont
  {{Dwyer}}, \citenamefont {{Edo}}, \citenamefont {{Edwards}}, \citenamefont
  {{Effler}}, \citenamefont {{Eggenstein}}, \citenamefont {{Ehrens}},
  \citenamefont {{Eichholz}}, \citenamefont {{Eikenberry}}, \citenamefont
  {{Engels}}, \citenamefont {{Essick}}, \citenamefont {{Etzel}}, \citenamefont
  {{Evans}}, \citenamefont {{Evans}}, \citenamefont {{Everett}}, \citenamefont
  {{Factourovich}}, \citenamefont {{Fafone}}, \citenamefont {{Fair}},
  \citenamefont {{Fairhurst}}, \citenamefont {{Fan}}, \citenamefont {{Fang}},
  \citenamefont {{Farinon}}, \citenamefont {{Farr}}, \citenamefont {{Farr}},
  \citenamefont {{Favata}}, \citenamefont {{Fays}}, \citenamefont {{Fehrmann}},
  \citenamefont {{Fejer}}, \citenamefont {{Ferrante}}, \citenamefont
  {{Ferreira}}, \citenamefont {{Ferrini}}, \citenamefont {{Fidecaro}},
  \citenamefont {{Fiori}}, \citenamefont {{Fiorucci}}, \citenamefont
  {{Fisher}}, \citenamefont {{Flaminio}}, \citenamefont {{Fletcher}},
  \citenamefont {{Fournier}}, \citenamefont {{Frasca}}, \citenamefont
  {{Frasconi}}, \citenamefont {{Frei}}, \citenamefont {{Freise}}, \citenamefont
  {{Frey}}, \citenamefont {{Frey}}, \citenamefont {{Fricke}}, \citenamefont
  {{Fritschel}}, \citenamefont {{Frolov}}, \citenamefont {{Fulda}},
  \citenamefont {{Fyffe}}, \citenamefont {{Gabbard}}, \citenamefont {{Gair}},
  \citenamefont {{Gammaitoni}}, \citenamefont {{Gaonkar}}, \citenamefont
  {{Garufi}}, \citenamefont {{Gaur}}, \citenamefont {{Gehrels}}, \citenamefont
  {{Gemme}}, \citenamefont {{Genin}}, \citenamefont {{Gennai}}, \citenamefont
  {{George}}, \citenamefont {{Gergely}}, \citenamefont {{Germain}},
  \citenamefont {{Ghosh}}, \citenamefont {{Ghosh}}, \citenamefont {{Giaime}},
  \citenamefont {{Giardina}}, \citenamefont {{Giazotto}}, \citenamefont
  {{Gill}}, \citenamefont {{Glaefke}}, \citenamefont {{Goetz}}, \citenamefont
  {{Goetz}}, \citenamefont {{Gondan}}, \citenamefont {{Gonz{\'a}lez}},
  \citenamefont {{Gonzalez Castro}}, \citenamefont {{Gopakumar}}, \citenamefont
  {{Gordon}}, \citenamefont {{Gorodetsky}}, \citenamefont {{Gossan}},
  \citenamefont {{Gosselin}}, \citenamefont {{Gouaty}}, \citenamefont
  {{Grado}}, \citenamefont {{Graef}}, \citenamefont {{Graff}}, \citenamefont
  {{Granata}}, \citenamefont {{Grant}}, \citenamefont {{Gras}}, \citenamefont
  {{Gray}}, \citenamefont {{Greco}}, \citenamefont {{Green}}, \citenamefont
  {{Groot}}, \citenamefont {{Grote}}, \citenamefont {{Grunewald}},
  \citenamefont {{Guidi}}, \citenamefont {{Guo}}, \citenamefont {{Gupta}},
  \citenamefont {{Gupta}}, \citenamefont {{Gushwa}}, \citenamefont
  {{Gustafson}}, \citenamefont {{Gustafson}}, \citenamefont {{Hacker}},
  \citenamefont {{Hall}}, \citenamefont {{Hall}}, \citenamefont {{Hammond}},
  \citenamefont {{Haney}}, \citenamefont {{Hanke}}, \citenamefont {{Hanks}},
  \citenamefont {{Hanna}}, \citenamefont {{Hannam}}, \citenamefont {{Hanson}},
  \citenamefont {{Hardwick}}, \citenamefont {{Harms}}, \citenamefont {{Harry}},
  \citenamefont {{Harry}}, \citenamefont {{Hart}}, \citenamefont {{Hartman}},
  \citenamefont {{Haster}}, \citenamefont {{Haughian}}, \citenamefont
  {{Heidmann}}, \citenamefont {{Heintze}}, \citenamefont {{Heitmann}},
  \citenamefont {{Hello}}, \citenamefont {{Hemming}}, \citenamefont {{Hendry}},
  \citenamefont {{Heng}}, \citenamefont {{Hennig}}, \citenamefont
  {{Heptonstall}}, \citenamefont {{Heurs}}, \citenamefont {{Hild}},
  \citenamefont {{Hoak}}, \citenamefont {{Hodge}}, \citenamefont {{Hofman}},
  \citenamefont {{Hollitt}}, \citenamefont {{Holt}}, \citenamefont {{Holz}},
  \citenamefont {{Hopkins}}, \citenamefont {{Hosken}}, \citenamefont {{Hough}},
  \citenamefont {{Houston}}, \citenamefont {{Howell}}, \citenamefont {{Hu}},
  \citenamefont {{Huang}}, \citenamefont {{Huerta}}, \citenamefont {{Huet}},
  \citenamefont {{Hughey}}, \citenamefont {{Husa}}, \citenamefont {{Huttner}},
  \citenamefont {{Huynh-Dinh}}, \citenamefont {{Idrisy}}, \citenamefont
  {{Indik}}, \citenamefont {{Ingram}}, \citenamefont {{Inta}}, \citenamefont
  {{Isa}}, \citenamefont {{Isac}}, \citenamefont {{Isi}}, \citenamefont
  {{Islas}}, \citenamefont {{Isogai}}, \citenamefont {{Iyer}}, \citenamefont
  {{Izumi}}, \citenamefont {{Jacqmin}}, \citenamefont {{Jang}}, \citenamefont
  {{Jani}}, \citenamefont {{Jaranowski}}, \citenamefont {{Jawahar}},
  \citenamefont {{Jim{\'e}nez-Forteza}}, \citenamefont {{Johnson}},
  \citenamefont {{Jones}}, \citenamefont {{Jones}}, \citenamefont {{Jonker}},
  \citenamefont {{Ju}}, \citenamefont {{Haris}}, \citenamefont {{Kalaghatgi}},
  \citenamefont {{Kalmus}}, \citenamefont {{Kalogera}}, \citenamefont
  {{Kamaretsos}}, \citenamefont {{Kandhasamy}}, \citenamefont {{Kang}},
  \citenamefont {{Kanner}}, \citenamefont {{Karki}}, \citenamefont
  {{Kasprzack}}, \citenamefont {{Katsavounidis}}, \citenamefont {{Katzman}},
  \citenamefont {{Kaufer}}, \citenamefont {{Kaur}}, \citenamefont {{Kawabe}},
  \citenamefont {{Kawazoe}}, \citenamefont {{K{\'e}f{\'e}lian}}, \citenamefont
  {{Kehl}}, \citenamefont {{Keitel}}, \citenamefont {{Kelley}}, \citenamefont
  {{Kells}}, \citenamefont {{Kennedy}}, \citenamefont {{Key}}, \citenamefont
  {{Khalaidovski}}, \citenamefont {{Khalili}}, \citenamefont {{Khan}},
  \citenamefont {{Khan}}, \citenamefont {{Khan}}, \citenamefont {{Khazanov}},
  \citenamefont {{Kijbunchoo}}, \citenamefont {{Kim}}, \citenamefont {{Kim}},
  \citenamefont {{Kim}}, \citenamefont {{Kim}}, \citenamefont {{Kim}},
  \citenamefont {{Kim}}, \citenamefont {{King}}, \citenamefont {{King}},
  \citenamefont {{Kinzel}}, \citenamefont {{Kissel}}, \citenamefont
  {{Kleybolte}}, \citenamefont {{Klimenko}}, \citenamefont {{Koehlenbeck}},
  \citenamefont {{Kokeyama}}, \citenamefont {{Koley}}, \citenamefont
  {{Kondrashov}}, \citenamefont {{Kontos}}, \citenamefont {{Korobko}},
  \citenamefont {{Korth}}, \citenamefont {{Kowalska}}, \citenamefont {{Kozak}},
  \citenamefont {{Kringel}}, \citenamefont {{Krishnan}}, \citenamefont
  {{Kr{\'o}lak}}, \citenamefont {{Krueger}}, \citenamefont {{Kuehn}},
  \citenamefont {{Kumar}}, \citenamefont {{Kuo}}, \citenamefont {{Kutynia}},
  \citenamefont {{Lackey}}, \citenamefont {{Landry}}, \citenamefont {{Lange}},
  \citenamefont {{Lantz}}, \citenamefont {{Lasky}}, \citenamefont
  {{Lazzarini}}, \citenamefont {{Lazzaro}}, \citenamefont {{Leaci}},
  \citenamefont {{Leavey}}, \citenamefont {{Lebigot}}, \citenamefont {{Lee}},
  \citenamefont {{Lee}}, \citenamefont {{Lee}}, \citenamefont {{Lee}},
  \citenamefont {{Lenon}}, \citenamefont {{Leonardi}}, \citenamefont {{Leong}},
  \citenamefont {{Leroy}}, \citenamefont {{Letendre}}, \citenamefont {{Levin}},
  \citenamefont {{Levine}}, \citenamefont {{Li}}, \citenamefont {{Libson}},
  \citenamefont {{Littenberg}}, \citenamefont {{Lockerbie}}, \citenamefont
  {{Loew}}, \citenamefont {{Logue}}, \citenamefont {{Lombardi}}, \citenamefont
  {{Lord}}, \citenamefont {{Lorenzini}}, \citenamefont {{Loriette}},
  \citenamefont {{Lormand}}, \citenamefont {{Losurdo}}, \citenamefont
  {{Lough}}, \citenamefont {{L{\"u}ck}}, \citenamefont {{Lundgren}},
  \citenamefont {{Luo}}, \citenamefont {{Lynch}}, \citenamefont {{Ma}},
  \citenamefont {{MacDonald}}, \citenamefont {{Machenschalk}}, \citenamefont
  {{MacInnis}}, \citenamefont {{Macleod}}, \citenamefont
  {{Maga{\~n}a-Sandoval}}, \citenamefont {{Magee}}, \citenamefont
  {{Mageswaran}}, \citenamefont {{Majorana}}, \citenamefont {{Maksimovic}},
  \citenamefont {{Malvezzi}}, \citenamefont {{Man}}, \citenamefont {{Mandel}},
  \citenamefont {{Mandic}}, \citenamefont {{Mangano}}, \citenamefont
  {{Mansell}}, \citenamefont {{Manske}}, \citenamefont {{Mantovani}},
  \citenamefont {{Marchesoni}}, \citenamefont {{Marion}}, \citenamefont
  {{M{\'a}rka}}, \citenamefont {{M{\'a}rka}}, \citenamefont {{Markosyan}},
  \citenamefont {{Maros}}, \citenamefont {{Martelli}}, \citenamefont
  {{Martellini}}, \citenamefont {{Martin}}, \citenamefont {{Martin}},
  \citenamefont {{Martynov}}, \citenamefont {{Marx}}, \citenamefont {{Mason}},
  \citenamefont {{Masserot}}, \citenamefont {{Massinger}}, \citenamefont
  {{Masso-Reid}}, \citenamefont {{Mastrogiovanni}}, \citenamefont
  {{Matichard}}, \citenamefont {{Matone}}, \citenamefont {{Mavalvala}},
  \citenamefont {{Mazumder}}, \citenamefont {{Mazzolo}}, \citenamefont
  {{McCarthy}}, \citenamefont {{McClelland }}, \citenamefont {{McCormick}},
  \citenamefont {{McGuire}}, \citenamefont {{McIntyre}}, \citenamefont
  {{McIver}}, \citenamefont {{McManus}}, \citenamefont {{McWilliams}},
  \citenamefont {{Meacher}}, \citenamefont {{Meadors}}, \citenamefont
  {{Meidam}}, \citenamefont {{Melatos}}, \citenamefont {{Mendell}},
  \citenamefont {{Mendoza-Gandara}}, \citenamefont {{Mercer}}, \citenamefont
  {{Merilh}}, \citenamefont {{Merzougui}}, \citenamefont {{Meshkov}},
  \citenamefont {{Messenger}}, \citenamefont {{Messick}}, \citenamefont
  {{Metzdorff}}, \citenamefont {{Meyers}}, \citenamefont {{Mezzani}},
  \citenamefont {{Miao}}, \citenamefont {{Michel}}, \citenamefont
  {{Middleton}}, \citenamefont {{Mikhailov}}, \citenamefont {{Milano}},
  \citenamefont {{Miller}}, \citenamefont {{Miller}}, \citenamefont
  {{Millhouse}}, \citenamefont {{Minenkov}}, \citenamefont {{Ming}},
  \citenamefont {{Mirshekari}}, \citenamefont {{Mishra}}, \citenamefont
  {{Mitra}}, \citenamefont {{Mitrofanov}}, \citenamefont {{Mitselmakher}},
  \citenamefont {{Mittleman}}, \citenamefont {{Moggi}}, \citenamefont
  {{Mohan}}, \citenamefont {{Mohapatra}}, \citenamefont {{Montani}},
  \citenamefont {{Moore}}, \citenamefont {{Moore}}, \citenamefont {{Moraru}},
  \citenamefont {{Moreno}}, \citenamefont {{Morriss}}, \citenamefont
  {{Mossavi}}, \citenamefont {{Mours}}, \citenamefont {{Mow-Lowry}},
  \citenamefont {{Mueller}}, \citenamefont {{Mueller}}, \citenamefont {{Muir}},
  \citenamefont {{Mukherjee}}, \citenamefont {{Mukherjee}}, \citenamefont
  {{Mukherjee}}, \citenamefont {{Mukund}}, \citenamefont {{Mullavey}},
  \citenamefont {{Munch}}, \citenamefont {{Murphy}}, \citenamefont {{Murray}},
  \citenamefont {{Mytidis}}, \citenamefont {{Nardecchia}}, \citenamefont
  {{Naticchioni}}, \citenamefont {{Nayak}}, \citenamefont {{Necula}},
  \citenamefont {{Nedkova}}, \citenamefont {{Nelemans}}, \citenamefont
  {{Neri}}, \citenamefont {{Neunzert}}, \citenamefont {{Newton}}, \citenamefont
  {{Nguyen}}, \citenamefont {{Nielsen}}, \citenamefont {{Nissanke}},
  \citenamefont {{Nitz}}, \citenamefont {{Nocera}}, \citenamefont {{Nolting}},
  \citenamefont {{Normandin}}, \citenamefont {{Nuttall}}, \citenamefont
  {{Oberling}}, \citenamefont {{Ochsner}}, \citenamefont {{O'Dell}},
  \citenamefont {{Oelker}}, \citenamefont {{Ogin}}, \citenamefont {{Oh}},
  \citenamefont {{Oh}}, \citenamefont {{Ohme}}, \citenamefont {{Oliver}},
  \citenamefont {{Oppermann}}, \citenamefont {{Oram}}, \citenamefont
  {{O'Reilly}}, \citenamefont {{O'Shaughnessy}}, \citenamefont {{Ott}},
  \citenamefont {{Ottaway}}, \citenamefont {{Ottens}}, \citenamefont
  {{Overmier}}, \citenamefont {{Owen}}, \citenamefont {{Pai}}, \citenamefont
  {{Pai}}, \citenamefont {{Palamos}}, \citenamefont {{Palashov}}, \citenamefont
  {{Palomba}}, \citenamefont {{Pal-Singh}}, \citenamefont {{Pan}},
  \citenamefont {{Pankow}}, \citenamefont {{Pannarale}}, \citenamefont
  {{Pant}}, \citenamefont {{Paoletti}}, \citenamefont {{Paoli}}, \citenamefont
  {{Papa}}, \citenamefont {{Paris}}, \citenamefont {{Parker}}, \citenamefont
  {{Pascucci}}, \citenamefont {{Pasqualetti}}, \citenamefont {{Passaquieti}},
  \citenamefont {{Passuello}}, \citenamefont {{Patricelli}}, \citenamefont
  {{Patrick}}, \citenamefont {{Pearlstone}}, \citenamefont {{Pedraza}},
  \citenamefont {{Pedurand}}, \citenamefont {{Pekowsky}}, \citenamefont
  {{Pele}}, \citenamefont {{Penn}}, \citenamefont {{Pereira}}, \citenamefont
  {{Perreca}}, \citenamefont {{Phelps}}, \citenamefont {{Piccinni}},
  \citenamefont {{Pichot}}, \citenamefont {{Piergiovanni}}, \citenamefont
  {{Pierro}}, \citenamefont {{Pillant}}, \citenamefont {{Pinard}},
  \citenamefont {{Pinto}}, \citenamefont {{Pitkin}}, \citenamefont
  {{Poggiani}}, \citenamefont {{Popolizio}}, \citenamefont {{Post}},
  \citenamefont {{Powell}}, \citenamefont {{Prasad}}, \citenamefont {{Predoi}},
  \citenamefont {{Premachand ra}}, \citenamefont {{Prestegard}}, \citenamefont
  {{Price}}, \citenamefont {{Prijatelj}}, \citenamefont {{Principe}},
  \citenamefont {{Privitera}}, \citenamefont {{Prix}}, \citenamefont {{Prodi}},
  \citenamefont {{Prokhorov}}, \citenamefont {{Puncken}}, \citenamefont
  {{Punturo}}, \citenamefont {{Puppo}}, \citenamefont {{P{\"u}rrer}},
  \citenamefont {{Qi}}, \citenamefont {{Qin}}, \citenamefont {{Quetschke}},
  \citenamefont {{Quintero}}, \citenamefont {{Quitzow-James}}, \citenamefont
  {{Raab}}, \citenamefont {{Rabeling}}, \citenamefont {{Radkins}},
  \citenamefont {{Raffai}}, \citenamefont {{Raja}}, \citenamefont
  {{Rakhmanov}}, \citenamefont {{Rapagnani}}, \citenamefont {{Raymond}},
  \citenamefont {{Razzano}}, \citenamefont {{Re}}, \citenamefont {{Read}},
  \citenamefont {{Reed}}, \citenamefont {{Regimbau}}, \citenamefont {{Rei}},
  \citenamefont {{Reid}}, \citenamefont {{Reitze}}, \citenamefont {{Rew}},
  \citenamefont {{Ricci}}, \citenamefont {{Riles}}, \citenamefont
  {{Robertson}}, \citenamefont {{Robie}}, \citenamefont {{Robinet}},
  \citenamefont {{Rocchi}}, \citenamefont {{Rolland}}, \citenamefont
  {{Rollins}}, \citenamefont {{Roma}}, \citenamefont {{Romano}}, \citenamefont
  {{Romano}}, \citenamefont {{Romanov}}, \citenamefont {{Romie}}, \citenamefont
  {{Rosi{\'n}ska}}, \citenamefont {{Rowan}}, \citenamefont {{R{\"u}diger}},
  \citenamefont {{Ruggi}}, \citenamefont {{Ryan}}, \citenamefont {{Sachdev}},
  \citenamefont {{Sadecki}}, \citenamefont {{Sadeghian}}, \citenamefont
  {{Salconi}}, \citenamefont {{Saleem}}, \citenamefont {{Salemi}},
  \citenamefont {{Samajdar}}, \citenamefont {{Sammut}}, \citenamefont
  {{Sanchez}}, \citenamefont {{Sand berg}}, \citenamefont {{Sandeen}},
  \citenamefont {{Sanders}}, \citenamefont {{Santamaria}}, \citenamefont
  {{Sassolas}}, \citenamefont {{Sathyaprakash}}, \citenamefont {{Saulson}},
  \citenamefont {{Sauter}}, \citenamefont {{Savage}}, \citenamefont
  {{Sawadsky}}, \citenamefont {{Schale}}, \citenamefont {{Schilling}},
  \citenamefont {{Schmidt}}, \citenamefont {{Schmidt}}, \citenamefont
  {{Schnabel}}, \citenamefont {{Schofield}}, \citenamefont {{Sch{\"o}nbeck}},
  \citenamefont {{Schreiber}}, \citenamefont {{Schuette}}, \citenamefont
  {{Schutz}}, \citenamefont {{Scott}}, \citenamefont {{Scott}}, \citenamefont
  {{Sellers}}, \citenamefont {{Sentenac}}, \citenamefont {{Sequino}},
  \citenamefont {{Sergeev}}, \citenamefont {{Serna}}, \citenamefont
  {{Setyawati}}, \citenamefont {{Sevigny}}, \citenamefont {{Shaddock}},
  \citenamefont {{Shahriar}}, \citenamefont {{Shaltev}}, \citenamefont
  {{Shao}}, \citenamefont {{Shapiro}}, \citenamefont {{Shawhan}}, \citenamefont
  {{Sheperd}}, \citenamefont {{Shoemaker}}, \citenamefont {{Shoemaker}},
  \citenamefont {{Siellez}}, \citenamefont {{Siemens}}, \citenamefont
  {{Sieniawska}}, \citenamefont {{Sigg}}, \citenamefont {{Silva}},
  \citenamefont {{Simakov}}, \citenamefont {{Singer}}, \citenamefont
  {{Singer}}, \citenamefont {{Singh}}, \citenamefont {{Singh}}, \citenamefont
  {{Singhal}}, \citenamefont {{Sintes}}, \citenamefont {{Slagmolen}},
  \citenamefont {{Smith}}, \citenamefont {{Smith}}, \citenamefont {{Smith}},
  \citenamefont {{Son}}, \citenamefont {{Sorazu}}, \citenamefont
  {{Sorrentino}}, \citenamefont {{Souradeep}}, \citenamefont {{Srivastava}},
  \citenamefont {{Staley}}, \citenamefont {{Steinke}}, \citenamefont
  {{Steinlechner}}, \citenamefont {{Steinlechner}}, \citenamefont
  {{Steinmeyer}}, \citenamefont {{Stephens}}, \citenamefont {{Stone}},
  \citenamefont {{Strain}}, \citenamefont {{Straniero}}, \citenamefont
  {{Stratta}}, \citenamefont {{Strauss}}, \citenamefont {{Strigin}},
  \citenamefont {{Sturani}}, \citenamefont {{Stuver}}, \citenamefont
  {{Summerscales}}, \citenamefont {{Sun}}, \citenamefont {{Sutton}},
  \citenamefont {{Swinkels}}, \citenamefont {{Szczepa{\'n}czyk}}, \citenamefont
  {{Tacca}}, \citenamefont {{Talukder}}, \citenamefont {{Tanner}},
  \citenamefont {{T{\'a}pai}}, \citenamefont {{Tarabrin}}, \citenamefont
  {{Taracchini}}, \citenamefont {{Taylor}}, \citenamefont {{Theeg}},
  \citenamefont {{Thirugnanasambandam}}, \citenamefont {{Thomas}},
  \citenamefont {{Thomas}}, \citenamefont {{Thomas}}, \citenamefont {{Thorne}},
  \citenamefont {{Thorne}}, \citenamefont {{Thrane}}, \citenamefont {{Tiwari}},
  \citenamefont {{Tiwari}}, \citenamefont {{Tokmakov}}, \citenamefont
  {{Tomlinson}}, \citenamefont {{Tonelli}}, \citenamefont {{Torres}},
  \citenamefont {{Torrie}}, \citenamefont {{T{\"o}yr{\"a}}}, \citenamefont
  {{Travasso}}, \citenamefont {{Traylor}}, \citenamefont {{Trifir{\`o}}},
  \citenamefont {{Tringali}}, \citenamefont {{Trozzo}}, \citenamefont {{Tse}},
  \citenamefont {{Turconi}}, \citenamefont {{Tuyenbayev}}, \citenamefont
  {{Ugolini}}, \citenamefont {{Unnikrishnan}}, \citenamefont {{Urban}},
  \citenamefont {{Usman}}, \citenamefont {{Vahlbruch}}, \citenamefont
  {{Vajente}}, \citenamefont {{Valdes}}, \citenamefont {{van Bakel}},
  \citenamefont {{van Beuzekom}}, \citenamefont {{van den Brand}},
  \citenamefont {{Van Den Broeck}}, \citenamefont {{Vander-Hyde}},
  \citenamefont {{van der Schaaf}}, \citenamefont {{van Heijningen}},
  \citenamefont {{van Veggel}}, \citenamefont {{Vardaro}}, \citenamefont
  {{Vass}}, \citenamefont {{Vas{\'u}th}}, \citenamefont {{Vaulin}},
  \citenamefont {{Vecchio}}, \citenamefont {{Vedovato}}, \citenamefont
  {{Veitch}}, \citenamefont {{Veitch}}, \citenamefont {{Venkateswara}},
  \citenamefont {{Verkindt}}, \citenamefont {{Vetrano}}, \citenamefont
  {{Vicer{\'e}}}, \citenamefont {{Vinciguerra}}, \citenamefont {{Vine}},
  \citenamefont {{Vinet}}, \citenamefont {{Vitale}}, \citenamefont {{Vo}},
  \citenamefont {{Vocca}}, \citenamefont {{Vorvick}}, \citenamefont {{Voss}},
  \citenamefont {{Vousden}}, \citenamefont {{Vyatchanin}}, \citenamefont
  {{Wade}}, \citenamefont {{Wade}}, \citenamefont {{Wade}}, \citenamefont
  {{Walker}}, \citenamefont {{Wallace}}, \citenamefont {{Walsh}}, \citenamefont
  {{Wang}}, \citenamefont {{Wang}}, \citenamefont {{Wang}}, \citenamefont
  {{Wang}}, \citenamefont {{Wang}}, \citenamefont {{Ward}}, \citenamefont
  {{Warner}}, \citenamefont {{Was}}, \citenamefont {{Weaver}}, \citenamefont
  {{Wei}}, \citenamefont {{Weinert}}, \citenamefont {{Weinstein}},
  \citenamefont {{Weiss}}, \citenamefont {{Welborn}}, \citenamefont {{Wen}},
  \citenamefont {{We{\ss}els}}, \citenamefont {{Westphal}}, \citenamefont
  {{Wette}}, \citenamefont {{Whelan}}, \citenamefont {{Whitcomb}},
  \citenamefont {{White}}, \citenamefont {{Whiting}}, \citenamefont
  {{Williams}}, \citenamefont {{Williamson}}, \citenamefont {{Willis}},
  \citenamefont {{Willke}}, \citenamefont {{Wimmer}}, \citenamefont
  {{Winkler}}, \citenamefont {{Wipf}}, \citenamefont {{Wittel}}, \citenamefont
  {{Woan}}, \citenamefont {{Worden}}, \citenamefont {{Wright}}, \citenamefont
  {{Wu}}, \citenamefont {{Yablon}}, \citenamefont {{Yam}}, \citenamefont
  {{Yamamoto}}, \citenamefont {{Yancey}}, \citenamefont {{Yap}}, \citenamefont
  {{Yu}}, \citenamefont {{Yvert}}, \citenamefont {{Zadro{\.Z}ny}},
  \citenamefont {{Zangrando}}, \citenamefont {{Zanolin}}, \citenamefont
  {{Zendri}}, \citenamefont {{Zevin}}, \citenamefont {{Zhang}}, \citenamefont
  {{Zhang}}, \citenamefont {{Zhang}}, \citenamefont {{Zhang}}, \citenamefont
  {{Zhao}}, \citenamefont {{Zhou}}, \citenamefont {{Zhou}}, \citenamefont
  {{Zhu}}, \citenamefont {{Zucker}}, \citenamefont {{Zuraw}}, \citenamefont
  {{Zweizig}}, \citenamefont {{LIGO Scientific Collaboration}},\ and\
  \citenamefont {{Virgo Collaboration}}}]{2016PhRvD..94j2001A}%
  \BibitemOpen
  \bibfield  {author} {\bibinfo {author} {B.~P. {Abbott}}, \bibinfo {author}
  {R.~{Abbott}}, \bibinfo {author} {T.~D. {Abbott}}, \bibinfo {author} {M.~R.
  {Abernathy}}, \bibinfo {author} {F.~{Acernese}}, \bibinfo {author}
  {K.~{Ackley}}, \bibinfo {author} {C.~{Adams}}, \bibinfo {author}
  {T.~{Adams}}, \bibinfo {author} {P.~{Addesso}}, \bibinfo {author} {R.~X.
  {Adhikari}}, et~al.,\ }\href {https://doi.org/10.1103/PhysRevD.94.102001}
  {\bibfield  {journal} {\bibinfo  {journal} {\prd}\ }\textbf {\bibinfo
  {volume} {94}},\ \bibinfo {eid} {102001} (\bibinfo {year} {2016})},\ \Eprint
  {https://arxiv.org/abs/1605.01785} {arXiv:1605.01785 [gr-qc]} \BibitemShut
  {NoStop}%
\bibitem [{\citenamefont {{Ott}}(2009)}]{2009CQGra..26t4015O}%
  \BibitemOpen
  \bibfield  {author} {\bibinfo {author} {C.~D. {Ott}},\ }\href
  {https://doi.org/10.1088/0264-9381/26/20/204015} {\bibfield  {journal}
  {\bibinfo  {journal} {Class Quantum Gravity}\ }\textbf {\bibinfo {volume}
  {26}},\ \bibinfo {eid} {204015} (\bibinfo {year} {2009})},\ \Eprint
  {https://arxiv.org/abs/0905.2797} {arXiv:0905.2797 [astro-ph.HE]}
  \BibitemShut {NoStop}%
\bibitem [{\citenamefont {{Dimmelmeier}}\ \emph {et~al.}(2007)\citenamefont
  {{Dimmelmeier}}, \citenamefont {{Ott}}, \citenamefont {{Janka}},
  \citenamefont {{Marek}},\ and\ \citenamefont
  {{M{\"u}ller}}}]{2007PhRvL..98y1101D}%
  \BibitemOpen
  \bibfield  {author} {\bibinfo {author} {H.~{Dimmelmeier}}, \bibinfo {author}
  {C.~D. {Ott}}, \bibinfo {author} {H.~T. {Janka}}, \bibinfo {author}
  {A.~{Marek}},\ and\ \bibinfo {author} {E.~{M{\"u}ller}},\ }\href
  {https://doi.org/10.1103/PhysRevLett.98.251101} {\bibfield  {journal}
  {\bibinfo  {journal} {\prl}\ }\textbf {\bibinfo {volume} {98}},\ \bibinfo
  {eid} {251101} (\bibinfo {year} {2007})},\ \Eprint
  {https://arxiv.org/abs/astro-ph/0702305} {arXiv:astro-ph/0702305 [astro-ph]}
  \BibitemShut {NoStop}%
\bibitem [{\citenamefont {{Fryer}}\ and\ \citenamefont
  {{New}}(2011)}]{2011LRR....14....1F}%
  \BibitemOpen
  \bibfield  {author} {\bibinfo {author} {C.~L. {Fryer}}\ and\ \bibinfo
  {author} {K.~C.~B. {New}},\ }\href {https://doi.org/10.12942/lrr-2011-1}
  {\bibfield  {journal} {\bibinfo  {journal} {Living Rev. Relativ.}\ }\textbf
  {\bibinfo {volume} {14}},\ \bibinfo {eid} {1} (\bibinfo {year}
  {2011})}\BibitemShut {NoStop}%
\bibitem [{\citenamefont {{Hayama}}\ \emph {et~al.}(2016)\citenamefont
  {{Hayama}}, \citenamefont {{Kuroda}}, \citenamefont {{Nakamura}},\ and\
  \citenamefont {{Yamada}}}]{2016PhRvL.116o1102H}%
  \BibitemOpen
  \bibfield  {author} {\bibinfo {author} {K.~{Hayama}}, \bibinfo {author}
  {T.~{Kuroda}}, \bibinfo {author} {K.~{Nakamura}},\ and\ \bibinfo {author}
  {S.~{Yamada}},\ }\href {https://doi.org/10.1103/PhysRevLett.116.151102}
  {\bibfield  {journal} {\bibinfo  {journal} {\prl}\ }\textbf {\bibinfo
  {volume} {116}},\ \bibinfo {eid} {151102} (\bibinfo {year} {2016})},\ \Eprint
  {https://arxiv.org/abs/1606.01520} {arXiv:1606.01520 [astro-ph.HE]}
  \BibitemShut {NoStop}%
\bibitem [{\citenamefont {{O'Connor}}\ and\ \citenamefont
  {{Couch}}(2018{\natexlab{a}})}]{2018ApJ...865...81O}%
  \BibitemOpen
  \bibfield  {author} {\bibinfo {author} {E.~P. {O'Connor}}\ and\ \bibinfo
  {author} {S.~M. {Couch}},\ }\href {https://doi.org/10.3847/1538-4357/aadcf7}
  {\bibfield  {journal} {\bibinfo  {journal} {\apj}\ }\textbf {\bibinfo
  {volume} {865}},\ \bibinfo {eid} {81} (\bibinfo {year}
  {2018}{\natexlab{a}})},\ \Eprint {https://arxiv.org/abs/1807.07579}
  {arXiv:1807.07579 [astro-ph.HE]} \BibitemShut {NoStop}%
\bibitem [{\citenamefont {{Radice}}\ \emph {et~al.}(2019)\citenamefont
  {{Radice}}, \citenamefont {{Morozova}}, \citenamefont {{Burrows}},
  \citenamefont {{Vartanyan}},\ and\ \citenamefont
  {{Nagakura}}}]{2019ApJ...876L...9R}%
  \BibitemOpen
  \bibfield  {author} {\bibinfo {author} {D.~{Radice}}, \bibinfo {author}
  {V.~{Morozova}}, \bibinfo {author} {A.~{Burrows}}, \bibinfo {author}
  {D.~{Vartanyan}},\ and\ \bibinfo {author} {H.~{Nagakura}},\ }\href
  {https://doi.org/10.3847/2041-8213/ab191a} {\bibfield  {journal} {\bibinfo
  {journal} {\apjl}\ }\textbf {\bibinfo {volume} {876}},\ \bibinfo {eid} {L9}
  (\bibinfo {year} {2019})},\ \Eprint {https://arxiv.org/abs/1812.07703}
  {arXiv:1812.07703 [astro-ph.HE]} \BibitemShut {NoStop}%
\bibitem [{\citenamefont {{Lin}}\ \emph {et~al.}(2006)\citenamefont {{Lin}},
  \citenamefont {{Cheng}}, \citenamefont {{Chu}},\ and\ \citenamefont
  {{Suen}}}]{2006ApJ...639..382L}%
  \BibitemOpen
  \bibfield  {author} {\bibinfo {author} {L.~M. {Lin}}, \bibinfo {author}
  {K.~S. {Cheng}}, \bibinfo {author} {M.~C. {Chu}},\ and\ \bibinfo {author}
  {W.~M. {Suen}},\ }\href {https://doi.org/10.1086/499202} {\bibfield
  {journal} {\bibinfo  {journal} {\apj}\ }\textbf {\bibinfo {volume} {639}},\
  \bibinfo {pages} {382} (\bibinfo {year} {2006})},\ \Eprint
  {https://arxiv.org/abs/astro-ph/0509447} {arXiv:astro-ph/0509447 [astro-ph]}
  \BibitemShut {NoStop}%
\bibitem [{\citenamefont {{Abdikamalov}}\ \emph {et~al.}(2009)\citenamefont
  {{Abdikamalov}}, \citenamefont {{Dimmelmeier}}, \citenamefont {{Rezzolla}},\
  and\ \citenamefont {{Miller}}}]{2009MNRAS.392...52A}%
  \BibitemOpen
  \bibfield  {author} {\bibinfo {author} {E.~B. {Abdikamalov}}, \bibinfo
  {author} {H.~{Dimmelmeier}}, \bibinfo {author} {L.~{Rezzolla}},\ and\
  \bibinfo {author} {J.~C. {Miller}},\ }\href
  {https://doi.org/10.1111/j.1365-2966.2008.14056.x} {\bibfield  {journal}
  {\bibinfo  {journal} {\mnras}\ }\textbf {\bibinfo {volume} {392}},\ \bibinfo
  {pages} {52} (\bibinfo {year} {2009})},\ \Eprint
  {https://arxiv.org/abs/0806.1700} {arXiv:0806.1700 [astro-ph]} \BibitemShut
  {NoStop}%
\bibitem [{\citenamefont {{Dimmelmeier}}\ \emph {et~al.}(2009)\citenamefont
  {{Dimmelmeier}}, \citenamefont {{Bejger}}, \citenamefont {{Haensel}},\ and\
  \citenamefont {{Zdunik}}}]{2009MNRAS.396.2269D}%
  \BibitemOpen
  \bibfield  {author} {\bibinfo {author} {H.~{Dimmelmeier}}, \bibinfo {author}
  {M.~{Bejger}}, \bibinfo {author} {P.~{Haensel}},\ and\ \bibinfo {author}
  {J.~L. {Zdunik}},\ }\href {https://doi.org/10.1111/j.1365-2966.2009.14891.x}
  {\bibfield  {journal} {\bibinfo  {journal} {\mnras}\ }\textbf {\bibinfo
  {volume} {396}},\ \bibinfo {pages} {2269} (\bibinfo {year} {2009})},\ \Eprint
  {https://arxiv.org/abs/0901.3819} {arXiv:0901.3819 [astro-ph.SR]}
  \BibitemShut {NoStop}%
\bibitem [{\citenamefont {{Sagert}}\ \emph {et~al.}(2010)\citenamefont
  {{Sagert}}, \citenamefont {{Fischer}}, \citenamefont {{Hempel}},
  \citenamefont {{Pagliara}}, \citenamefont {{Schaffner-Bielich}},
  \citenamefont {{Thielemann}},\ and\ \citenamefont
  {{Liebend{\"o}rfer}}}]{2010JPhG...37i4064S}%
  \BibitemOpen
  \bibfield  {author} {\bibinfo {author} {I.~{Sagert}}, \bibinfo {author}
  {T.~{Fischer}}, \bibinfo {author} {M.~{Hempel}}, \bibinfo {author}
  {G.~{Pagliara}}, \bibinfo {author} {J.~{Schaffner-Bielich}}, \bibinfo
  {author} {F.~K. {Thielemann}},\ and\ \bibinfo {author}
  {M.~{Liebend{\"o}rfer}},\ }\href
  {https://doi.org/10.1088/0954-3899/37/9/094064} {\bibfield  {journal}
  {\bibinfo  {journal} {J. Phys. G. Nucl. Part. Phys.}\ }\textbf {\bibinfo
  {volume} {37}},\ \bibinfo {eid} {094064} (\bibinfo {year} {2010})},\ \Eprint
  {https://arxiv.org/abs/1003.2320} {arXiv:1003.2320 [astro-ph.HE]}
  \BibitemShut {NoStop}%
\bibitem [{\citenamefont {compose.obspm.fr/eos/75/}()}]{compose}%
  \BibitemOpen
  \bibfield  {author} {\bibinfo {author} {compose.obspm.fr/eos/75/},\
  }\href@noop {} {}\BibitemShut {NoStop}%
\bibitem [{\citenamefont {{Shen}}\ \emph {et~al.}(1998)\citenamefont {{Shen}},
  \citenamefont {{Toki}}, \citenamefont {{Oyamatsu}},\ and\ \citenamefont
  {{Sumiyoshi}}}]{1998PThPh.100.1013S}%
  \BibitemOpen
  \bibfield  {author} {\bibinfo {author} {H.~{Shen}}, \bibinfo {author}
  {H.~{Toki}}, \bibinfo {author} {K.~{Oyamatsu}},\ and\ \bibinfo {author}
  {K.~{Sumiyoshi}},\ }\href {https://doi.org/10.1143/PTP.100.1013} {\bibfield
  {journal} {\bibinfo  {journal} {Progress of Theoretical Physics}\ }\textbf
  {\bibinfo {volume} {100}},\ \bibinfo {pages} {1013} (\bibinfo {year}
  {1998})},\ \Eprint {https://arxiv.org/abs/nucl-th/9806095}
  {arXiv:nucl-th/9806095 [nucl-th]} \BibitemShut {NoStop}%
\bibitem [{\citenamefont {{Glendenning}}(1992)}]{1992PhRvD..46.1274G}%
  \BibitemOpen
  \bibfield  {author} {\bibinfo {author} {N.~K. {Glendenning}},\ }\href
  {https://doi.org/10.1103/PhysRevD.46.1274} {\bibfield  {journal} {\bibinfo
  {journal} {\prd}\ }\textbf {\bibinfo {volume} {46}},\ \bibinfo {pages} {1274}
  (\bibinfo {year} {1992})}\BibitemShut {NoStop}%
\bibitem [{\citenamefont {{Hempel}}\ \emph {et~al.}(2016)\citenamefont
  {{Hempel}}, \citenamefont {{Heinimann}}, \citenamefont {{Yudin}},
  \citenamefont {{Iosilevskiy}}, \citenamefont {{Liebend{\"o}rfer}},\ and\
  \citenamefont {{Thielemann}}}]{2016PhRvD..94j3001H}%
  \BibitemOpen
  \bibfield  {author} {\bibinfo {author} {M.~{Hempel}}, \bibinfo {author}
  {O.~{Heinimann}}, \bibinfo {author} {A.~{Yudin}}, \bibinfo {author}
  {I.~{Iosilevskiy}}, \bibinfo {author} {M.~{Liebend{\"o}rfer}},\ and\ \bibinfo
  {author} {F.-K. {Thielemann}},\ }\href
  {https://doi.org/10.1103/PhysRevD.94.103001} {\bibfield  {journal} {\bibinfo
  {journal} {\prd}\ }\textbf {\bibinfo {volume} {94}},\ \bibinfo {eid} {103001}
  (\bibinfo {year} {2016})},\ \Eprint {https://arxiv.org/abs/1511.06551}
  {arXiv:1511.06551 [nucl-th]} \BibitemShut {NoStop}%
\bibitem [{\citenamefont {{Fryxell}}\ \emph {et~al.}(2000)\citenamefont
  {{Fryxell}}, \citenamefont {{Olson}}, \citenamefont {{Ricker}}, \citenamefont
  {{Timmes}}, \citenamefont {{Zingale}}, \citenamefont {{Lamb}}, \citenamefont
  {{MacNeice}}, \citenamefont {{Rosner}}, \citenamefont {{Truran}},\ and\
  \citenamefont {{Tufo}}}]{2000ApJS..131..273F}%
  \BibitemOpen
  \bibfield  {author} {\bibinfo {author} {B.~{Fryxell}}, \bibinfo {author}
  {K.~{Olson}}, \bibinfo {author} {P.~{Ricker}}, \bibinfo {author} {F.~X.
  {Timmes}}, \bibinfo {author} {M.~{Zingale}}, \bibinfo {author} {D.~Q.
  {Lamb}}, \bibinfo {author} {P.~{MacNeice}}, \bibinfo {author} {R.~{Rosner}},
  \bibinfo {author} {J.~W. {Truran}},\ and\ \bibinfo {author} {H.~{Tufo}},\
  }\href {https://doi.org/10.1086/317361} {\bibfield  {journal} {\bibinfo
  {journal} {\apjs}\ }\textbf {\bibinfo {volume} {131}},\ \bibinfo {pages}
  {273} (\bibinfo {year} {2000})}\BibitemShut {NoStop}%
\bibitem [{\citenamefont {{O'Connor}}\ and\ \citenamefont
  {{Couch}}(2018{\natexlab{b}})}]{2018ApJ...854...63O}%
  \BibitemOpen
  \bibfield  {author} {\bibinfo {author} {E.~P. {O'Connor}}\ and\ \bibinfo
  {author} {S.~M. {Couch}},\ }\href {https://doi.org/10.3847/1538-4357/aaa893}
  {\bibfield  {journal} {\bibinfo  {journal} {\apj}\ }\textbf {\bibinfo
  {volume} {854}},\ \bibinfo {eid} {63} (\bibinfo {year}
  {2018}{\natexlab{b}})},\ \Eprint {https://arxiv.org/abs/1511.07443}
  {arXiv:1511.07443 [astro-ph.HE]} \BibitemShut {NoStop}%
\bibitem [{\citenamefont {{Woosley}}\ \emph {et~al.}(2002)\citenamefont
  {{Woosley}}, \citenamefont {{Heger}},\ and\ \citenamefont
  {{Weaver}}}]{2002RvMP...74.1015W}%
  \BibitemOpen
  \bibfield  {author} {\bibinfo {author} {S.~E. {Woosley}}, \bibinfo {author}
  {A.~{Heger}},\ and\ \bibinfo {author} {T.~A. {Weaver}},\ }\href
  {https://doi.org/10.1103/RevModPhys.74.1015} {\bibfield  {journal} {\bibinfo
  {journal} {Rev. Mod. Phys.}\ }\textbf {\bibinfo {volume} {74}},\ \bibinfo
  {pages} {1015} (\bibinfo {year} {2002})}\BibitemShut {NoStop}%
\bibitem [{\citenamefont {{Marek}}\ \emph {et~al.}(2006)\citenamefont
  {{Marek}}, \citenamefont {{Dimmelmeier}}, \citenamefont {{Janka}},
  \citenamefont {{M{\"u}ller}},\ and\ \citenamefont
  {{Buras}}}]{2006A&A...445..273M}%
  \BibitemOpen
  \bibfield  {author} {\bibinfo {author} {A.~{Marek}}, \bibinfo {author}
  {H.~{Dimmelmeier}}, \bibinfo {author} {H.~T. {Janka}}, \bibinfo {author}
  {E.~{M{\"u}ller}},\ and\ \bibinfo {author} {R.~{Buras}},\ }\href
  {https://doi.org/10.1051/0004-6361:20052840} {\bibfield  {journal} {\bibinfo
  {journal} {\aap}\ }\textbf {\bibinfo {volume} {445}},\ \bibinfo {pages} {273}
  (\bibinfo {year} {2006})},\ \Eprint {https://arxiv.org/abs/astro-ph/0502161}
  {arXiv:astro-ph/0502161 [astro-ph]} \BibitemShut {NoStop}%
\bibitem [{SI()}]{SI}%
  \BibitemOpen
  \href@noop {} {\bibinfo {title} {See supplementary materials for the gr
  approximations, test of potential numerical artifacts, resolution dependence,
  spatial distribution of $h_+$, and the gw spectrogram for the stos eos, which
  includes refs. \cite{2019ApJ...878...13P,2010CQGra..27k4103O}}}\BibitemShut
  {NoStop}%
\bibitem [{\citenamefont {{Obergaulinger}}\ and\ \citenamefont
  {{Aloy}}(2020)}]{2020MNRAS.492.4613O}%
  \BibitemOpen
  \bibfield  {author} {\bibinfo {author} {M.~{Obergaulinger}}\ and\ \bibinfo
  {author} {M.~{\'A}. {Aloy}},\ }\href {https://doi.org/10.1093/mnras/staa096}
  {\bibfield  {journal} {\bibinfo  {journal} {\mnras}\ }\textbf {\bibinfo
  {volume} {492}},\ \bibinfo {pages} {4613} (\bibinfo {year} {2020})},\ \Eprint
  {https://arxiv.org/abs/1909.01105} {arXiv:1909.01105 [astro-ph.HE]}
  \BibitemShut {NoStop}%
\bibitem [{\citenamefont {{Finn}}\ and\ \citenamefont
  {{Evans}}(1990)}]{1990ApJ...351..588F}%
  \BibitemOpen
  \bibfield  {author} {\bibinfo {author} {L.~S. {Finn}}\ and\ \bibinfo {author}
  {C.~R. {Evans}},\ }\href {https://doi.org/10.1086/168497} {\bibfield
  {journal} {\bibinfo  {journal} {\apj}\ }\textbf {\bibinfo {volume} {351}},\
  \bibinfo {pages} {588} (\bibinfo {year} {1990})}\BibitemShut {NoStop}%
\bibitem [{\citenamefont {{Torres-Forn{\'e}}}\ \emph
  {et~al.}(2019)\citenamefont {{Torres-Forn{\'e}}}, \citenamefont
  {{Cerd{\'a}-Dur{\'a}n}}, \citenamefont {{Obergaulinger}}, \citenamefont
  {{M{\"u}ller}},\ and\ \citenamefont {{Font}}}]{2019PhRvL.123e1102T}%
  \BibitemOpen
  \bibfield  {author} {\bibinfo {author} {A.~{Torres-Forn{\'e}}}, \bibinfo
  {author} {P.~{Cerd{\'a}-Dur{\'a}n}}, \bibinfo {author} {M.~{Obergaulinger}},
  \bibinfo {author} {B.~{M{\"u}ller}},\ and\ \bibinfo {author} {J.~A. {Font}},\
  }\href {https://doi.org/10.1103/PhysRevLett.123.051102} {\bibfield  {journal}
  {\bibinfo  {journal} {\prl}\ }\textbf {\bibinfo {volume} {123}},\ \bibinfo
  {eid} {051102} (\bibinfo {year} {2019})},\ \Eprint
  {https://arxiv.org/abs/1902.10048} {arXiv:1902.10048 [gr-qc]} \BibitemShut
  {NoStop}%
\bibitem [{\citenamefont {{M{\"u}ller}}\ \emph {et~al.}(2013)\citenamefont
  {{M{\"u}ller}}, \citenamefont {{Janka}},\ and\ \citenamefont
  {{Marek}}}]{2013ApJ...766...43M}%
  \BibitemOpen
  \bibfield  {author} {\bibinfo {author} {B.~{M{\"u}ller}}, \bibinfo {author}
  {H.-T. {Janka}},\ and\ \bibinfo {author} {A.~{Marek}},\ }\href
  {https://doi.org/10.1088/0004-637X/766/1/43} {\bibfield  {journal} {\bibinfo
  {journal} {\apj}\ }\textbf {\bibinfo {volume} {766}},\ \bibinfo {eid} {43}
  (\bibinfo {year} {2013})},\ \Eprint {https://arxiv.org/abs/1210.6984}
  {arXiv:1210.6984 [astro-ph.SR]} \BibitemShut {NoStop}%
\bibitem [{\citenamefont {{Flanagan}}\ and\ \citenamefont
  {{Hughes}}(1998)}]{1998PhRvD..57.4535F}%
  \BibitemOpen
  \bibfield  {author} {\bibinfo {author} {{\'E}.~{\'E}. {Flanagan}}\ and\
  \bibinfo {author} {S.~A. {Hughes}},\ }\href
  {https://doi.org/10.1103/PhysRevD.57.4535} {\bibfield  {journal} {\bibinfo
  {journal} {\prd}\ }\textbf {\bibinfo {volume} {57}},\ \bibinfo {pages} {4535}
  (\bibinfo {year} {1998})},\ \Eprint {https://arxiv.org/abs/gr-qc/9701039}
  {arXiv:gr-qc/9701039 [gr-qc]} \BibitemShut {NoStop}%
\bibitem [{\citenamefont {{Punturo}}\ \emph {et~al.}(2010)\citenamefont
  {{Punturo}}, \citenamefont {{Abernathy}}, \citenamefont {{Acernese}},
  \citenamefont {{Allen}}, \citenamefont {{Andersson}}, \citenamefont {{Arun}},
  \citenamefont {{Barone}}, \citenamefont {{Barr}}, \citenamefont
  {{Barsuglia}}, \citenamefont {{Beker}}, \citenamefont {{Beveridge}},
  \citenamefont {{Birindelli}}, \citenamefont {{Bose}}, \citenamefont {{Bosi}},
  \citenamefont {{Braccini}}, \citenamefont {{Bradaschia}}, \citenamefont
  {{Bulik}}, \citenamefont {{Calloni}}, \citenamefont {{Cella}}, \citenamefont
  {{Chassande Mottin}}, \citenamefont {{Chelkowski}}, \citenamefont
  {{Chincarini}}, \citenamefont {{Clark}}, \citenamefont {{Coccia}},
  \citenamefont {{Colacino}}, \citenamefont {{Colas}}, \citenamefont
  {{Cumming}}, \citenamefont {{Cunningham}}, \citenamefont {{Cuoco}},
  \citenamefont {{Danilishin}}, \citenamefont {{Danzmann}}, \citenamefont {{De
  Luca}}, \citenamefont {{De Salvo}}, \citenamefont {{Dent}}, \citenamefont
  {{De Rosa}}, \citenamefont {{Di Fiore}}, \citenamefont {{Di Virgilio}},
  \citenamefont {{Doets}}, \citenamefont {{Fafone}}, \citenamefont {{Falferi}},
  \citenamefont {{Flaminio}}, \citenamefont {{Franc}}, \citenamefont
  {{Frasconi}}, \citenamefont {{Freise}}, \citenamefont {{Fulda}},
  \citenamefont {{Gair}}, \citenamefont {{Gemme}}, \citenamefont {{Gennai}},
  \citenamefont {{Giazotto}}, \citenamefont {{Glampedakis}}, \citenamefont
  {{Granata}}, \citenamefont {{Grote}}, \citenamefont {{Guidi}}, \citenamefont
  {{Hammond}}, \citenamefont {{Hannam}}, \citenamefont {{Harms}}, \citenamefont
  {{Heinert}}, \citenamefont {{Hendry}}, \citenamefont {{Heng}}, \citenamefont
  {{Hennes}}, \citenamefont {{Hild}}, \citenamefont {{Hough}}, \citenamefont
  {{Husa}}, \citenamefont {{Huttner}}, \citenamefont {{Jones}}, \citenamefont
  {{Khalili}}, \citenamefont {{Kokeyama}}, \citenamefont {{Kokkotas}},
  \citenamefont {{Krishnan}}, \citenamefont {{Lorenzini}}, \citenamefont
  {{L{\"u}ck}}, \citenamefont {{Majorana}}, \citenamefont {{Mandel}},
  \citenamefont {{Mandic}}, \citenamefont {{Martin}}, \citenamefont {{Michel}},
  \citenamefont {{Minenkov}}, \citenamefont {{Morgado}}, \citenamefont
  {{Mosca}}, \citenamefont {{Mours}}, \citenamefont {{M{\"u}ller─Ebhardt}},
  \citenamefont {{Murray}}, \citenamefont {{Nawrodt}}, \citenamefont
  {{Nelson}}, \citenamefont {{Oshaughnessy}}, \citenamefont {{Ott}},
  \citenamefont {{Palomba}}, \citenamefont {{Paoli}}, \citenamefont
  {{Parguez}}, \citenamefont {{Pasqualetti}}, \citenamefont {{Passaquieti}},
  \citenamefont {{Passuello}}, \citenamefont {{Pinard}}, \citenamefont
  {{Poggiani}}, \citenamefont {{Popolizio}}, \citenamefont {{Prato}},
  \citenamefont {{Puppo}}, \citenamefont {{Rabeling}}, \citenamefont
  {{Rapagnani}}, \citenamefont {{Read}}, \citenamefont {{Regimbau}},
  \citenamefont {{Rehbein}}, \citenamefont {{Reid}}, \citenamefont
  {{Rezzolla}}, \citenamefont {{Ricci}}, \citenamefont {{Richard}},
  \citenamefont {{Rocchi}}, \citenamefont {{Rowan}}, \citenamefont
  {{R{\"u}diger}}, \citenamefont {{Sassolas}}, \citenamefont {{Sathyaprakash}},
  \citenamefont {{Schnabel}}, \citenamefont {{Schwarz}}, \citenamefont
  {{Seidel}}, \citenamefont {{Sintes}}, \citenamefont {{Somiya}}, \citenamefont
  {{Speirits}}, \citenamefont {{Strain}}, \citenamefont {{Strigin}},
  \citenamefont {{Sutton}}, \citenamefont {{Tarabrin}}, \citenamefont
  {{Th{\"u}ring}}, \citenamefont {{van den Brand}}, \citenamefont {{van
  Leewen}}, \citenamefont {{van Veggel}}, \citenamefont {{van den Broeck}},
  \citenamefont {{Vecchio}}, \citenamefont {{Veitch}}, \citenamefont
  {{Vetrano}}, \citenamefont {{Vicere}}, \citenamefont {{Vyatchanin}},
  \citenamefont {{Willke}}, \citenamefont {{Woan}}, \citenamefont
  {{Wolfango}},\ and\ \citenamefont {{Yamamoto}}}]{2010CQGra..27s4002P}%
  \BibitemOpen
  \bibfield  {author} {\bibinfo {author} {M.~{Punturo}}, \bibinfo {author}
  {M.~{Abernathy}}, \bibinfo {author} {F.~{Acernese}}, \bibinfo {author}
  {B.~{Allen}}, \bibinfo {author} {N.~{Andersson}}, \bibinfo {author}
  {K.~{Arun}}, \bibinfo {author} {F.~{Barone}}, \bibinfo {author} {B.~{Barr}},
  \bibinfo {author} {M.~{Barsuglia}}, \bibinfo {author} {M.~{Beker}}, et~al.,\
  }\href {https://doi.org/10.1088/0264-9381/27/19/194002} {\bibfield  {journal}
  {\bibinfo  {journal} {Classical and Quantum Gravity}\ }\textbf {\bibinfo
  {volume} {27}},\ \bibinfo {eid} {194002} (\bibinfo {year}
  {2010})}\BibitemShut {NoStop}%
\bibitem [{\citenamefont {{Abbott}}\ \emph {et~al.}(2017)\citenamefont
  {{Abbott}}, \citenamefont {{Abbott}}, \citenamefont {{Abbott}}, \citenamefont
  {{Abernathy}}, \citenamefont {{Ackley}}, \citenamefont {{Adams}},
  \citenamefont {{Addesso}}, \citenamefont {{Adhikari}}, \citenamefont
  {{Adya}}, \citenamefont {{Affeldt}}, \citenamefont {{Aggarwal}},
  \citenamefont {{Aguiar}}, \citenamefont {{Ain}}, \citenamefont {{Ajith}},
  \citenamefont {{Allen}}, \citenamefont {{Altin}}, \citenamefont {{Anderson}},
  \citenamefont {{Anderson}}, \citenamefont {{Arai}}, \citenamefont {{Araya}},
  \citenamefont {{Arceneaux}}, \citenamefont {{Areeda}}, \citenamefont
  {{Arun}}, \citenamefont {{Ashton}}, \citenamefont {{Ast}}, \citenamefont
  {{Aston}}, \citenamefont {{Aufmuth}}, \citenamefont {{Aulbert}},
  \citenamefont {{Babak}}, \citenamefont {{Baker}}, \citenamefont {{Ballmer}},
  \citenamefont {{Barayoga}}, \citenamefont {{Barclay}}, \citenamefont
  {{Barish}}, \citenamefont {{Barker}}, \citenamefont {{Barr}}, \citenamefont
  {{Barsotti}}, \citenamefont {{Bartlett}}, \citenamefont {{Bartos}},
  \citenamefont {{Bassiri}}, \citenamefont {{Batch}}, \citenamefont {{Baune}},
  \citenamefont {{Bell}}, \citenamefont {{Berger}}, \citenamefont {{Bergmann}},
  \citenamefont {{Berry}}, \citenamefont {{Betzwieser}}, \citenamefont
  {{Bhagwat}}, \citenamefont {{Bhand are}}, \citenamefont {{Bilenko}},
  \citenamefont {{Billingsley}}, \citenamefont {{Birch}}, \citenamefont
  {{Birney}}, \citenamefont {{Biscans}}, \citenamefont {{Bisht}}, \citenamefont
  {{Biwer}}, \citenamefont {{Blackburn}}, \citenamefont {{Blair}},
  \citenamefont {{Blair}}, \citenamefont {{Blair}}, \citenamefont {{Bock}},
  \citenamefont {{Bogan}}, \citenamefont {{Bohe}}, \citenamefont {{Bond}},
  \citenamefont {{Bork}}, \citenamefont {{Bose}}, \citenamefont {{Brady}},
  \citenamefont {{Braginsky}}, \citenamefont {{Brau}}, \citenamefont
  {{Brinkmann}}, \citenamefont {{Brockill}}, \citenamefont {{Broida}},
  \citenamefont {{Brooks}}, \citenamefont {{Brown}}, \citenamefont {{Brown}},
  \citenamefont {{Brown}}, \citenamefont {{Brunett}}, \citenamefont
  {{Buchanan}}, \citenamefont {{Buikema}}, \citenamefont {{Buonanno}},
  \citenamefont {{Byer}}, \citenamefont {{Cabero}}, \citenamefont {{Cadonati}},
  \citenamefont {{Cahillane}}, \citenamefont {{Calder{\'o}n Bustillo}},
  \citenamefont {{Callister}}, \citenamefont {{Camp}}, \citenamefont
  {{Cannon}}, \citenamefont {{Cao}}, \citenamefont {{Capano}}, \citenamefont
  {{Caride}}, \citenamefont {{Caudill}}, \citenamefont {{Cavagli{\`a}}},
  \citenamefont {{Cepeda}}, \citenamefont {{Chamberlin}}, \citenamefont
  {{Chan}}, \citenamefont {{Chao}}, \citenamefont {{Charlton}}, \citenamefont
  {{Cheeseboro}}, \citenamefont {{Chen}}, \citenamefont {{Chen}}, \citenamefont
  {{Cheng}}, \citenamefont {{Cho}}, \citenamefont {{Cho}}, \citenamefont
  {{Chow}}, \citenamefont {{Christensen}}, \citenamefont {{Chu}}, \citenamefont
  {{Chung}}, \citenamefont {{Ciani}}, \citenamefont {{Clara}}, \citenamefont
  {{Clark}}, \citenamefont {{Collette}}, \citenamefont {{Cominsky}},
  \citenamefont {{Constancio}}, \citenamefont {{Cook}}, \citenamefont
  {{Corbitt}}, \citenamefont {{Cornish}}, \citenamefont {{Corsi}},
  \citenamefont {{Costa}}, \citenamefont {{Coughlin}}, \citenamefont
  {{Coughlin}}, \citenamefont {{Countryman}}, \citenamefont {{Couvares}},
  \citenamefont {{Cowan}}, \citenamefont {{Coward}}, \citenamefont {{Cowart}},
  \citenamefont {{Coyne}}, \citenamefont {{Coyne}}, \citenamefont {{Craig}},
  \citenamefont {{Creighton}}, \citenamefont {{Cripe}}, \citenamefont
  {{Crowder}}, \citenamefont {{Cumming}}, \citenamefont {{Cunningham}},
  \citenamefont {{Dal Canton}}, \citenamefont {{Danilishin}}, \citenamefont
  {{Danzmann}}, \citenamefont {{Darman}}, \citenamefont {{Dasgupta}},
  \citenamefont {{Da Silva Costa}}, \citenamefont {{Dave}}, \citenamefont
  {{Davies}}, \citenamefont {{Daw}}, \citenamefont {{De}}, \citenamefont
  {{DeBra}}, \citenamefont {{Del Pozzo}}, \citenamefont {{Denker}},
  \citenamefont {{Dent}}, \citenamefont {{Dergachev}}, \citenamefont
  {{DeRosa}}, \citenamefont {{DeSalvo}}, \citenamefont {{Devine}},
  \citenamefont {{Dhurand har}}, \citenamefont {{D{\'\i}az}}, \citenamefont
  {{Di Palma}}, \citenamefont {{Donovan}}, \citenamefont {{Dooley}},
  \citenamefont {{Doravari}}, \citenamefont {{Douglas}}, \citenamefont
  {{Downes}}, \citenamefont {{Drago}}, \citenamefont {{Drever}}, \citenamefont
  {{Driggers}}, \citenamefont {{Dwyer}}, \citenamefont {{Edo}}, \citenamefont
  {{Edwards}}, \citenamefont {{Effler}}, \citenamefont {{Eggenstein}},
  \citenamefont {{Ehrens}}, \citenamefont {{Eichholz}}, \citenamefont
  {{Eikenberry}}, \citenamefont {{Engels}}, \citenamefont {{Essick}},
  \citenamefont {{Etzel}}, \citenamefont {{Evans}}, \citenamefont {{Evans}},
  \citenamefont {{Everett}}, \citenamefont {{Factourovich}}, \citenamefont
  {{Fair}}, \citenamefont {{Fairhurst}}, \citenamefont {{Fan}}, \citenamefont
  {{Fang}}, \citenamefont {{Farr}}, \citenamefont {{Farr}}, \citenamefont
  {{Favata}}, \citenamefont {{Fays}}, \citenamefont {{Fehrmann}}, \citenamefont
  {{Fejer}}, \citenamefont {{Fenyvesi}}, \citenamefont {{Ferreira}},
  \citenamefont {{Fisher}}, \citenamefont {{Fletcher}}, \citenamefont {{Frei}},
  \citenamefont {{Freise}}, \citenamefont {{Frey}}, \citenamefont
  {{Fritschel}}, \citenamefont {{Frolov}}, \citenamefont {{Fulda}},
  \citenamefont {{Fyffe}}, \citenamefont {{Gabbard}}, \citenamefont {{Gair}},
  \citenamefont {{Gaonkar}}, \citenamefont {{Gaur}}, \citenamefont {{Gehrels}},
  \citenamefont {{Geng}}, \citenamefont {{George}}, \citenamefont {{Gergely}},
  \citenamefont {{Ghosh}}, \citenamefont {{Ghosh}}, \citenamefont {{Giaime}},
  \citenamefont {{Giardina}}, \citenamefont {{Gill}}, \citenamefont
  {{Glaefke}}, \citenamefont {{Goetz}}, \citenamefont {{Goetz}}, \citenamefont
  {{Gondan}}, \citenamefont {{Gonz{\'a}lez}}, \citenamefont {{Gopakumar}},
  \citenamefont {{Gordon}}, \citenamefont {{Gorodetsky}}, \citenamefont
  {{Gossan}}, \citenamefont {{Graef}}, \citenamefont {{Graff}}, \citenamefont
  {{Grant}}, \citenamefont {{Gras}}, \citenamefont {{Gray}}, \citenamefont
  {{Green}}, \citenamefont {{Grote}}, \citenamefont {{Grunewald}},
  \citenamefont {{Guo}}, \citenamefont {{Gupta}}, \citenamefont {{Gupta}},
  \citenamefont {{Gushwa}}, \citenamefont {{Gustafson}}, \citenamefont
  {{Gustafson}}, \citenamefont {{Hacker}}, \citenamefont {{Hall}},
  \citenamefont {{Hall}}, \citenamefont {{Hammond}}, \citenamefont {{Haney}},
  \citenamefont {{Hanke}}, \citenamefont {{Hanks}}, \citenamefont {{Hanna}},
  \citenamefont {{Hannam}}, \citenamefont {{Hanson}}, \citenamefont
  {{Hardwick}}, \citenamefont {{Harry}}, \citenamefont {{Harry}}, \citenamefont
  {{Hart}}, \citenamefont {{Hartman}}, \citenamefont {{Haster}}, \citenamefont
  {{Haughian}}, \citenamefont {{Heintze}}, \citenamefont {{Hendry}},
  \citenamefont {{Heng}}, \citenamefont {{Hennig}}, \citenamefont {{Henry}},
  \citenamefont {{Heptonstall}}, \citenamefont {{Heurs}}, \citenamefont
  {{Hild}}, \citenamefont {{Hoak}}, \citenamefont {{Holt}}, \citenamefont
  {{Holz}}, \citenamefont {{Hopkins}}, \citenamefont {{Hough}}, \citenamefont
  {{Houston}}, \citenamefont {{Howell}}, \citenamefont {{Hu}}, \citenamefont
  {{Huang}}, \citenamefont {{Huerta}}, \citenamefont {{Hughey}}, \citenamefont
  {{Husa}}, \citenamefont {{Huttner}}, \citenamefont {{Huynh-Dinh}},
  \citenamefont {{Indik}}, \citenamefont {{Ingram}}, \citenamefont {{Inta}},
  \citenamefont {{Isa}}, \citenamefont {{Isi}}, \citenamefont {{Isogai}},
  \citenamefont {{Iyer}}, \citenamefont {{Izumi}}, \citenamefont {{Jang}},
  \citenamefont {{Jani}}, \citenamefont {{Jawahar}}, \citenamefont {{Jian}},
  \citenamefont {{Jim{\'e}nez-Forteza}}, \citenamefont {{Johnson}},
  \citenamefont {{Jones}}, \citenamefont {{Jones}}, \citenamefont {{Ju}},
  \citenamefont {{Haris}}, \citenamefont {{Kalaghatgi}}, \citenamefont
  {{Kalogera}}, \citenamefont {{Kandhasamy}}, \citenamefont {{Kang}},
  \citenamefont {{Kanner}}, \citenamefont {{Kapadia}}, \citenamefont {{Karki}},
  \citenamefont {{Karvinen}}, \citenamefont {{Kasprzack}}, \citenamefont
  {{Katsavounidis}}, \citenamefont {{Katzman}}, \citenamefont {{Kaufer}},
  \citenamefont {{Kaur}}, \citenamefont {{Kawabe}}, \citenamefont {{Kehl}},
  \citenamefont {{Keitel}}, \citenamefont {{Kelley}}, \citenamefont {{Kells}},
  \citenamefont {{Kennedy}}, \citenamefont {{Key}}, \citenamefont {{Khalili}},
  \citenamefont {{Khan}}, \citenamefont {{Khan}}, \citenamefont {{Khazanov}},
  \citenamefont {{Kijbunchoo}}, \citenamefont {{Kim}}, \citenamefont {{Kim}},
  \citenamefont {{Kim}}, \citenamefont {{Kim}}, \citenamefont {{Kim}},
  \citenamefont {{Kim}}, \citenamefont {{Kim}}, \citenamefont {{Kimbrell}},
  \citenamefont {{King}}, \citenamefont {{King}}, \citenamefont {{Kissel}},
  \citenamefont {{Klein}}, \citenamefont {{Kleybolte}}, \citenamefont
  {{Klimenko}}, \citenamefont {{Koehlenbeck}}, \citenamefont {{Kondrashov}},
  \citenamefont {{Kontos}}, \citenamefont {{Korobko}}, \citenamefont {{Korth}},
  \citenamefont {{Kozak}}, \citenamefont {{Kringel}}, \citenamefont
  {{Krueger}}, \citenamefont {{Kuehn}}, \citenamefont {{Kumar}}, \citenamefont
  {{Kumar}}, \citenamefont {{Kuo}}, \citenamefont {{Lackey}}, \citenamefont
  {{Landry}}, \citenamefont {{Lange}}, \citenamefont {{Lantz}}, \citenamefont
  {{Lasky}}, \citenamefont {{Laxen}}, \citenamefont {{Lazzarini}},
  \citenamefont {{Leavey}}, \citenamefont {{Lebigot}}, \citenamefont {{Lee}},
  \citenamefont {{Lee}}, \citenamefont {{Lee}}, \citenamefont {{Lee}},
  \citenamefont {{Lenon}}, \citenamefont {{Leong}}, \citenamefont {{Levin}},
  \citenamefont {{Lewis}}, \citenamefont {{Li}}, \citenamefont {{Libson}},
  \citenamefont {{Littenberg}}, \citenamefont {{Lockerbie}}, \citenamefont
  {{Lombardi}}, \citenamefont {{London}}, \citenamefont {{Lord}}, \citenamefont
  {{Lormand }}, \citenamefont {{Lough}}, \citenamefont {{L{\"u}ck}},
  \citenamefont {{Lundgren}}, \citenamefont {{Lynch}}, \citenamefont {{Ma}},
  \citenamefont {{Machenschalk}}, \citenamefont {{MacInnis}}, \citenamefont
  {{Macleod}}, \citenamefont {{Maga{\~n}a-Sandoval}}, \citenamefont
  {{Maga{\~n}a Zertuche}}, \citenamefont {{Magee}}, \citenamefont {{Mandic}},
  \citenamefont {{Mangano}}, \citenamefont {{Mansell}}, \citenamefont
  {{Manske}}, \citenamefont {{M{\'a}rka}}, \citenamefont {{M{\'a}rka}},
  \citenamefont {{Markosyan}}, \citenamefont {{Maros}}, \citenamefont
  {{Martin}}, \citenamefont {{Martynov}}, \citenamefont {{Mason}},
  \citenamefont {{Massinger}}, \citenamefont {{Masso-Reid}}, \citenamefont
  {{Matichard}}, \citenamefont {{Matone}}, \citenamefont {{Mavalvala}},
  \citenamefont {{Mazumder}}, \citenamefont {{McCarthy}}, \citenamefont
  {{McClelland}}, \citenamefont {{McCormick}}, \citenamefont {{McGuire}},
  \citenamefont {{McIntyre}}, \citenamefont {{McIver}}, \citenamefont
  {{McManus}}, \citenamefont {{McRae}}, \citenamefont {{McWilliams}},
  \citenamefont {{Meacher}}, \citenamefont {{Meadors}}, \citenamefont
  {{Melatos}}, \citenamefont {{Mendell}}, \citenamefont {{Mercer}},
  \citenamefont {{Merilh}}, \citenamefont {{Meshkov}}, \citenamefont
  {{Messenger}}, \citenamefont {{Messick}}, \citenamefont {{Meyers}},
  \citenamefont {{Miao}}, \citenamefont {{Middleton}}, \citenamefont
  {{Mikhailov}}, \citenamefont {{Miller}}, \citenamefont {{Miller}},
  \citenamefont {{Miller}}, \citenamefont {{Miller}}, \citenamefont
  {{Millhouse}}, \citenamefont {{Ming}}, \citenamefont {{Mirshekari}},
  \citenamefont {{Mishra}}, \citenamefont {{Mitra}}, \citenamefont
  {{Mitrofanov}}, \citenamefont {{Mitselmakher}}, \citenamefont {{Mittleman}},
  \citenamefont {{Mohapatra}}, \citenamefont {{Moore}}, \citenamefont
  {{Moore}}, \citenamefont {{Moraru}}, \citenamefont {{Moreno}}, \citenamefont
  {{Morriss}}, \citenamefont {{Mossavi}}, \citenamefont {{Mow-Lowry}},
  \citenamefont {{Mueller}}, \citenamefont {{Muir}}, \citenamefont
  {{Mukherjee}}, \citenamefont {{Mukherjee}}, \citenamefont {{Mukherjee}},
  \citenamefont {{Mukund}}, \citenamefont {{Mullavey}}, \citenamefont
  {{Munch}}, \citenamefont {{Murphy}}, \citenamefont {{Murray}}, \citenamefont
  {{Mytidis}}, \citenamefont {{Nayak}}, \citenamefont {{Nedkova}},
  \citenamefont {{Nelson}}, \citenamefont {{Neunzert}}, \citenamefont
  {{Newton}}, \citenamefont {{Nguyen}}, \citenamefont {{Nielsen}},
  \citenamefont {{Nitz}}, \citenamefont {{Nolting}}, \citenamefont
  {{Normandin}}, \citenamefont {{Nuttall}}, \citenamefont {{Oberling}},
  \citenamefont {{Ochsner}}, \citenamefont {{O'Dell}}, \citenamefont
  {{Oelker}}, \citenamefont {{Ogin}}, \citenamefont {{Oh}}, \citenamefont
  {{Oh}}, \citenamefont {{Ohme}}, \citenamefont {{Oliver}}, \citenamefont
  {{Oppermann}}, \citenamefont {{Oram}}, \citenamefont {{O'Reilly}},
  \citenamefont {{O'Shaughnessy}}, \citenamefont {{Ottaway}}, \citenamefont
  {{Overmier}}, \citenamefont {{Owen}}, \citenamefont {{Pai}}, \citenamefont
  {{Pai}}, \citenamefont {{Palamos}}, \citenamefont {{Palashov}}, \citenamefont
  {{Pal-Singh}}, \citenamefont {{Pan}}, \citenamefont {{Pankow}}, \citenamefont
  {{Pannarale}}, \citenamefont {{Pant}}, \citenamefont {{Papa}}, \citenamefont
  {{Paris}}, \citenamefont {{Parker}}, \citenamefont {{Pascucci}},
  \citenamefont {{Patrick}}, \citenamefont {{Pearlstone}}, \citenamefont
  {{Pedraza}}, \citenamefont {{Pekowsky}}, \citenamefont {{Pele}},
  \citenamefont {{Penn}}, \citenamefont {{Perreca}}, \citenamefont {{Perri}},
  \citenamefont {{Phelps}}, \citenamefont {{Pierro}}, \citenamefont {{Pinto}},
  \citenamefont {{Pitkin}}, \citenamefont {{Poe}}, \citenamefont {{Post}},
  \citenamefont {{Powell}}, \citenamefont {{Prasad}}, \citenamefont {{Predoi}},
  \citenamefont {{Prestegard}}, \citenamefont {{Price}}, \citenamefont
  {{Prijatelj}}, \citenamefont {{Principe}}, \citenamefont {{Privitera}},
  \citenamefont {{Prokhorov}}, \citenamefont {{Puncken}}, \citenamefont
  {{P{\"u}rrer}}, \citenamefont {{Qi}}, \citenamefont {{Qin}}, \citenamefont
  {{Qiu}}, \citenamefont {{Quetschke}}, \citenamefont {{Quintero}},
  \citenamefont {{Quitzow-James}}, \citenamefont {{Raab}}, \citenamefont
  {{Rabeling}}, \citenamefont {{Radkins}}, \citenamefont {{Raffai}},
  \citenamefont {{Raja}}, \citenamefont {{Rajan}}, \citenamefont {{Rakhmanov}},
  \citenamefont {{Raymond}}, \citenamefont {{Read}}, \citenamefont {{Reed}},
  \citenamefont {{Reid}}, \citenamefont {{Reitze}}, \citenamefont {{Rew}},
  \citenamefont {{Reyes}}, \citenamefont {{Riles}}, \citenamefont {{Rizzo}},
  \citenamefont {{Robertson}}, \citenamefont {{Robie}}, \citenamefont
  {{Rollins}}, \citenamefont {{Roma}}, \citenamefont {{Romanov}}, \citenamefont
  {{Romie}}, \citenamefont {{Rowan}}, \citenamefont {{R{\"u}diger}},
  \citenamefont {{Ryan}}, \citenamefont {{Sachdev}}, \citenamefont {{Sadecki}},
  \citenamefont {{Sadeghian}}, \citenamefont {{Sakellariadou}}, \citenamefont
  {{Saleem}}, \citenamefont {{Salemi}}, \citenamefont {{Samajdar}},
  \citenamefont {{Sammut}}, \citenamefont {{Sanchez}}, \citenamefont {{Sand
  berg}}, \citenamefont {{Sandeen}}, \citenamefont {{Sanders}}, \citenamefont
  {{Sathyaprakash}}, \citenamefont {{Saulson}}, \citenamefont {{Sauter}},
  \citenamefont {{Savage}}, \citenamefont {{Sawadsky}}, \citenamefont
  {{Schale}}, \citenamefont {{Schilling}}, \citenamefont {{Schmidt}},
  \citenamefont {{Schmidt}}, \citenamefont {{Schnabel}}, \citenamefont
  {{Schofield}}, \citenamefont {{Sch{\"o}nbeck}}, \citenamefont {{Schreiber}},
  \citenamefont {{Schuette}}, \citenamefont {{Schutz}}, \citenamefont
  {{Scott}}, \citenamefont {{Scott}}, \citenamefont {{Sellers}}, \citenamefont
  {{Sengupta}}, \citenamefont {{Sergeev}}, \citenamefont {{Shaddock}},
  \citenamefont {{Shaffer}}, \citenamefont {{Shahriar}}, \citenamefont
  {{Shaltev}}, \citenamefont {{Shapiro}}, \citenamefont {{Shawhan}},
  \citenamefont {{Sheperd}}, \citenamefont {{Shoemaker}}, \citenamefont
  {{Shoemaker}}, \citenamefont {{Siellez}}, \citenamefont {{Siemens}},
  \citenamefont {{Sigg}}, \citenamefont {{Silva}}, \citenamefont {{Singer}},
  \citenamefont {{Singer}}, \citenamefont {{Singh}}, \citenamefont {{Singh}},
  \citenamefont {{Sintes}}, \citenamefont {{Slagmolen}}, \citenamefont
  {{Smith}}, \citenamefont {{Smith}}, \citenamefont {{Smith}}, \citenamefont
  {{Son}}, \citenamefont {{Sorazu}}, \citenamefont {{Souradeep}}, \citenamefont
  {{Srivastava}}, \citenamefont {{Staley}}, \citenamefont {{Steinke}},
  \citenamefont {{Steinlechner}}, \citenamefont {{Steinlechner}}, \citenamefont
  {{Steinmeyer}}, \citenamefont {{Stephens}}, \citenamefont {{Stone}},
  \citenamefont {{Strain}}, \citenamefont {{Strauss}}, \citenamefont
  {{Strigin}}, \citenamefont {{Sturani}}, \citenamefont {{Stuver}},
  \citenamefont {{Summerscales}}, \citenamefont {{Sun}}, \citenamefont
  {{Sunil}}, \citenamefont {{Sutton}}, \citenamefont {{Szczepa{\'n}czyk}},
  \citenamefont {{Talukder}}, \citenamefont {{Tanner}}, \citenamefont
  {{T{\'a}pai}}, \citenamefont {{Tarabrin}}, \citenamefont {{Taracchini}},
  \citenamefont {{Taylor}}, \citenamefont {{Theeg}}, \citenamefont
  {{Thirugnanasambandam}}, \citenamefont {{Thomas}}, \citenamefont {{Thomas}},
  \citenamefont {{Thomas}}, \citenamefont {{Thorne}}, \citenamefont {{Thrane}},
  \citenamefont {{Tiwari}}, \citenamefont {{Tokmakov}}, \citenamefont
  {{Toland}}, \citenamefont {{Tomlinson}}, \citenamefont {{Tornasi}},
  \citenamefont {{Torres}}, \citenamefont {{Torrie}}, \citenamefont
  {{T{\"o}yr{\"a}}}, \citenamefont {{Traylor}}, \citenamefont {{Trifir{\`o}}},
  \citenamefont {{Tse}}, \citenamefont {{Tuyenbayev}}, \citenamefont
  {{Ugolini}}, \citenamefont {{Unnikrishnan}}, \citenamefont {{Urban}},
  \citenamefont {{Usman}}, \citenamefont {{Vahlbruch}}, \citenamefont
  {{Vajente}}, \citenamefont {{Valdes}}, \citenamefont {{Vand er-Hyde}},
  \citenamefont {{van Veggel}}, \citenamefont {{Vass}}, \citenamefont
  {{Vaulin}}, \citenamefont {{Vecchio}}, \citenamefont {{Veitch}},
  \citenamefont {{Veitch}}, \citenamefont {{Venkateswara}}, \citenamefont
  {{Vinciguerra}}, \citenamefont {{Vine}}, \citenamefont {{Vitale}},
  \citenamefont {{Vo}}, \citenamefont {{Vorvick}}, \citenamefont {{Voss}},
  \citenamefont {{Vousden}}, \citenamefont {{Vyatchanin}}, \citenamefont
  {{Wade}}, \citenamefont {{Wade}}, \citenamefont {{Wade}}, \citenamefont
  {{Walker}}, \citenamefont {{Wallace}}, \citenamefont {{Walsh}}, \citenamefont
  {{Wang}}, \citenamefont {{Wang}}, \citenamefont {{Wang}}, \citenamefont
  {{Wang}}, \citenamefont {{Ward}}, \citenamefont {{Warner}}, \citenamefont
  {{Weaver}}, \citenamefont {{Weinert}}, \citenamefont {{Weinstein}},
  \citenamefont {{Weiss}}, \citenamefont {{Wen}}, \citenamefont {{We{\ss}els}},
  \citenamefont {{Westphal}}, \citenamefont {{Wette}}, \citenamefont
  {{Whelan}}, \citenamefont {{Whiting}}, \citenamefont {{Williams}},
  \citenamefont {{Williamson}}, \citenamefont {{Willis}}, \citenamefont
  {{Willke}}, \citenamefont {{Wimmer}}, \citenamefont {{Winkler}},
  \citenamefont {{Wipf}}, \citenamefont {{Wittel}}, \citenamefont {{Woan}},
  \citenamefont {{Woehler}}, \citenamefont {{Worden}}, \citenamefont
  {{Wright}}, \citenamefont {{Wu}}, \citenamefont {{Wu}}, \citenamefont
  {{Yablon}}, \citenamefont {{Yam}}, \citenamefont {{Yamamoto}}, \citenamefont
  {{Yancey}}, \citenamefont {{Yu}}, \citenamefont {{Zanolin}}, \citenamefont
  {{Zevin}}, \citenamefont {{Zhang}}, \citenamefont {{Zhang}}, \citenamefont
  {{Zhang}}, \citenamefont {{Zhao}}, \citenamefont {{Zhou}}, \citenamefont
  {{Zhou}}, \citenamefont {{Zhu}}, \citenamefont {{Zucker}}, \citenamefont
  {{Zuraw}}, \citenamefont {{Zweizig}}, \citenamefont {{(LIGO Scientific
  Collaboration}},\ and\ \citenamefont {{Harms}}}]{2017CQGra..34d4001A}%
  \BibitemOpen
  \bibfield  {author} {\bibinfo {author} {B.~P. {Abbott}}, \bibinfo {author}
  {R.~{Abbott}}, \bibinfo {author} {T.~D. {Abbott}}, \bibinfo {author} {M.~R.
  {Abernathy}}, \bibinfo {author} {K.~{Ackley}}, \bibinfo {author}
  {C.~{Adams}}, \bibinfo {author} {P.~{Addesso}}, \bibinfo {author} {R.~X.
  {Adhikari}}, \bibinfo {author} {V.~B. {Adya}}, \bibinfo {author}
  {C.~{Affeldt}}, et~al.,\ }\href {https://doi.org/10.1088/1361-6382/aa51f4}
  {\bibfield  {journal} {\bibinfo  {journal} {Classical and Quantum Gravity}\
  }\textbf {\bibinfo {volume} {34}},\ \bibinfo {eid} {044001} (\bibinfo {year}
  {2017})},\ \Eprint {https://arxiv.org/abs/1607.08697} {arXiv:1607.08697
  [astro-ph.IM]} \BibitemShut {NoStop}%
\bibitem [{\citenamefont {{Pan}}\ \emph {et~al.}(2018)\citenamefont {{Pan}},
  \citenamefont {{Liebend{\"o}rfer}}, \citenamefont {{Couch}},\ and\
  \citenamefont {{Thielemann}}}]{2018ApJ...857...13P}%
  \BibitemOpen
  \bibfield  {author} {\bibinfo {author} {K.-C. {Pan}}, \bibinfo {author}
  {M.~{Liebend{\"o}rfer}}, \bibinfo {author} {S.~M. {Couch}},\ and\ \bibinfo
  {author} {F.-K. {Thielemann}},\ }\href
  {https://doi.org/10.3847/1538-4357/aab71d} {\bibfield  {journal} {\bibinfo
  {journal} {\apj}\ }\textbf {\bibinfo {volume} {857}},\ \bibinfo {eid} {13}
  (\bibinfo {year} {2018})},\ \Eprint {https://arxiv.org/abs/1710.01690}
  {arXiv:1710.01690 [astro-ph.HE]} \BibitemShut {NoStop}%
\bibitem [{\citenamefont {{Barsotti}}\ \emph {et~al.}(2018)\citenamefont
  {{Barsotti}}, \citenamefont {{Fritschel}}, \citenamefont {{Evans}},\ and\
  \citenamefont {{Gras}}}]{LIGO_v5}%
  \BibitemOpen
  \bibfield  {author} {\bibinfo {author} {L.~{Barsotti}}, \bibinfo {author}
  {P.~{Fritschel}}, \bibinfo {author} {M.~{Evans}},\ and\ \bibinfo {author}
  {S.~{Gras}},\ }\href {dcc.ligo.org/LIGO-T1800042/public} {}\bibinfo {type}
  {Tech. Rep.}\ \bibinfo {number} {{LIGO-T1800042-v5}}\ (\bibinfo {year}
  {2018})\BibitemShut {NoStop}%
\bibitem [{\citenamefont {{Andresen}}\ \emph {et~al.}(2017)\citenamefont
  {{Andresen}}, \citenamefont {{M{\"u}ller}}, \citenamefont {{M{\"u}ller}},\
  and\ \citenamefont {{Janka}}}]{2017MNRAS.468.2032A}%
  \BibitemOpen
  \bibfield  {author} {\bibinfo {author} {H.~{Andresen}}, \bibinfo {author}
  {B.~{M{\"u}ller}}, \bibinfo {author} {E.~{M{\"u}ller}},\ and\ \bibinfo
  {author} {H.~T. {Janka}},\ }\href {https://doi.org/10.1093/mnras/stx618}
  {\bibfield  {journal} {\bibinfo  {journal} {\mnras}\ }\textbf {\bibinfo
  {volume} {468}},\ \bibinfo {pages} {2032} (\bibinfo {year} {2017})},\ \Eprint
  {https://arxiv.org/abs/1607.05199} {arXiv:1607.05199 [astro-ph.HE]}
  \BibitemShut {NoStop}%
\bibitem [{\citenamefont {{O'Connor}}\ and\ \citenamefont
  {{Ott}}(2011)}]{2011ApJ...730...70O}%
  \BibitemOpen
  \bibfield  {author} {\bibinfo {author} {E.~{O'Connor}}\ and\ \bibinfo
  {author} {C.~D. {Ott}},\ }\href {https://doi.org/10.1088/0004-637X/730/2/70}
  {\bibfield  {journal} {\bibinfo  {journal} {\apj}\ }\textbf {\bibinfo
  {volume} {730}},\ \bibinfo {eid} {70} (\bibinfo {year} {2011})},\ \Eprint
  {https://arxiv.org/abs/1010.5550} {arXiv:1010.5550 [astro-ph.HE]}
  \BibitemShut {NoStop}%
\bibitem [{\citenamefont {{Schneider}}\ \emph {et~al.}(2020)\citenamefont
  {{Schneider}}, \citenamefont {{O'Connor}}, \citenamefont {{Granqvist}},
  \citenamefont {{Betranhandy}},\ and\ \citenamefont
  {{Couch}}}]{2020ApJ...894....4S}%
  \BibitemOpen
  \bibfield  {author} {\bibinfo {author} {A.~d.~S. {Schneider}}, \bibinfo
  {author} {E.~{O'Connor}}, \bibinfo {author} {E.~{Granqvist}}, \bibinfo
  {author} {A.~{Betranhandy}},\ and\ \bibinfo {author} {S.~M. {Couch}},\ }\href
  {https://doi.org/10.3847/1538-4357/ab8308} {\bibfield  {journal} {\bibinfo
  {journal} {\apj}\ }\textbf {\bibinfo {volume} {894}},\ \bibinfo {eid} {4}
  (\bibinfo {year} {2020})},\ \Eprint {https://arxiv.org/abs/2001.10434}
  {arXiv:2001.10434 [astro-ph.HE]} \BibitemShut {NoStop}%
\bibitem [{\citenamefont {{Klimenko}}\ \emph {et~al.}(2008)\citenamefont
  {{Klimenko}}, \citenamefont {{Yakushin}}, \citenamefont {{Mercer}},\ and\
  \citenamefont {{Mitselmakher}}}]{2008CQGra..25k4029K}%
  \BibitemOpen
  \bibfield  {author} {\bibinfo {author} {S.~{Klimenko}}, \bibinfo {author}
  {I.~{Yakushin}}, \bibinfo {author} {A.~{Mercer}},\ and\ \bibinfo {author}
  {G.~{Mitselmakher}},\ }\href {https://doi.org/10.1088/0264-9381/25/11/114029}
  {\bibfield  {journal} {\bibinfo  {journal} {Class Quantum Gravity}\ }\textbf
  {\bibinfo {volume} {25}},\ \bibinfo {eid} {114029} (\bibinfo {year}
  {2008})},\ \Eprint {https://arxiv.org/abs/0802.3232} {arXiv:0802.3232
  [gr-qc]} \BibitemShut {NoStop}%
\bibitem [{\citenamefont {{Antoniadis}}\ \emph {et~al.}(2013)\citenamefont
  {{Antoniadis}}, \citenamefont {{Freire}}, \citenamefont {{Wex}},
  \citenamefont {{Tauris}}, \citenamefont {{Lynch}}, \citenamefont {{van
  Kerkwijk}}, \citenamefont {{Kramer}}, \citenamefont {{Bassa}}, \citenamefont
  {{Dhillon}}, \citenamefont {{Driebe}}, \citenamefont {{Hessels}},
  \citenamefont {{Kaspi}}, \citenamefont {{Kondratiev}}, \citenamefont
  {{Langer}}, \citenamefont {{Marsh}}, \citenamefont {{McLaughlin}},
  \citenamefont {{Pennucci}}, \citenamefont {{Ransom}}, \citenamefont
  {{Stairs}}, \citenamefont {{van Leeuwen}}, \citenamefont {{Verbiest}},\ and\
  \citenamefont {{Whelan}}}]{2013Sci...340..448A}%
  \BibitemOpen
  \bibfield  {author} {\bibinfo {author} {J.~{Antoniadis}}, \bibinfo {author}
  {P.~C.~C. {Freire}}, \bibinfo {author} {N.~{Wex}}, \bibinfo {author} {T.~M.
  {Tauris}}, \bibinfo {author} {R.~S. {Lynch}}, \bibinfo {author} {M.~H. {van
  Kerkwijk}}, \bibinfo {author} {M.~{Kramer}}, \bibinfo {author} {C.~{Bassa}},
  \bibinfo {author} {V.~S. {Dhillon}}, \bibinfo {author} {T.~{Driebe}},
  et~al.,\ }\href {https://doi.org/10.1126/science.1233232} {\bibfield
  {journal} {\bibinfo  {journal} {Science}\ }\textbf {\bibinfo {volume}
  {340}},\ \bibinfo {pages} {448} (\bibinfo {year} {2013})},\ \Eprint
  {https://arxiv.org/abs/1304.6875} {arXiv:1304.6875 [astro-ph.HE]}
  \BibitemShut {NoStop}%
\bibitem [{\citenamefont {{Pajkos}}\ \emph {et~al.}(2019)\citenamefont
  {{Pajkos}}, \citenamefont {{Couch}}, \citenamefont {{Pan}},\ and\
  \citenamefont {{O'Connor}}}]{2019ApJ...878...13P}%
  \BibitemOpen
  \bibfield  {author} {\bibinfo {author} {M.~A. {Pajkos}}, \bibinfo {author}
  {S.~M. {Couch}}, \bibinfo {author} {K.-C. {Pan}},\ and\ \bibinfo {author}
  {E.~P. {O'Connor}},\ }\href {https://doi.org/10.3847/1538-4357/ab1de2}
  {\bibfield  {journal} {\bibinfo  {journal} {\apj}\ }\textbf {\bibinfo
  {volume} {878}},\ \bibinfo {eid} {13} (\bibinfo {year} {2019})},\ \Eprint
  {https://arxiv.org/abs/1901.09055} {arXiv:1901.09055 [astro-ph.HE]}
  \BibitemShut {NoStop}%
\bibitem [{\citenamefont {{O'Connor}}\ and\ \citenamefont
  {{Ott}}(2010)}]{2010CQGra..27k4103O}%
  \BibitemOpen
  \bibfield  {author} {\bibinfo {author} {E.~{O'Connor}}\ and\ \bibinfo
  {author} {C.~D. {Ott}},\ }\href
  {https://doi.org/10.1088/0264-9381/27/11/114103} {\bibfield  {journal}
  {\bibinfo  {journal} {Class Quantum Gravity}\ }\textbf {\bibinfo {volume}
  {27}},\ \bibinfo {eid} {114103} (\bibinfo {year} {2010})},\ \Eprint
  {https://arxiv.org/abs/0912.2393} {arXiv:0912.2393 [astro-ph.HE]}
  \BibitemShut {NoStop}%
\end{thebibliography}%
	
	\clearpage
	\appendix
	\onecolumngrid
	\begin{center}
		\Large \textbf{Supplementary Materials: Gravitational-wave Signature of a First-order Quantum Chromodynamics Phase Transition in Core-Collapse Supernovae}
	\end{center}
	
\section{General relativistic approximations}
The FLASH code solves the Newtonian Euler equations. To mimic the deeper gravitational well in general relativity (GR), we use the effective gravitational potential with the Case A formalism of \cite{2006A&A...445..273M}, which has been tested in core-collapse supernova (CCSN) simulations routinely \cite{2018ApJ...854...63O,2019ApJ...878...13P,2020ApJ...894....4S}.

In this work, we extend this by including the GR time-dilation effect directly in the Euler equations. The modified Euler equations read:
\begin{equation}
	\label{eq:hydro} 
	\begin{aligned}
		\partial_t \rho + \nabla \cdot (\alpha \rho \vec{v})  &= 0,  \\
		\partial_t (\rho \vec{v}) + \nabla\cdot [\alpha(\rho\vec{v}\vec{v}+P)]  & = -\alpha (\rho-P/c^2) \nabla \Phi, \\
		\partial_t \tau + \nabla \cdot [\alpha(\tau+P)\vec{v}] & = -\alpha\rho\vec{v}\cdot \nabla \Phi, 
	\end{aligned}
\end{equation}
where $\rho$, $\vec{v}$, $P$ and $\tau$ are the rest mass density, velocity, pressure and kinetic plus internal energy density of the fluid; $\Phi$ is the effective GR gravitational potential; $\alpha=\exp(\Phi)$ is the lapse function. The lapse function is included in the fluxes and source terms. The additional source term in the momentum equation, $\alpha P/c^2\nabla \Phi$, is verified by the derivation of the GR hydrodynamic equations with the metric $g_{\mu\nu}=[-\alpha^2,1,r^2,r^2\sin\theta]$ in spherical symmetry \cite{2010CQGra..27k4103O}. This source term is important for maintaining the mechanical equilibrium in the hydrostatic regions. The energy equation is derived through the same procedure and has no additional source term. The consistency of the equations can be checked with a polytropic equation of state in which the pressure and specific internal energy are analytic functions of the rest mass density.

\begin{figure}[h!]
	\setcounter{figure}{0}
	\renewcommand{\thefigure}{S\arabic{figure}}
	\includegraphics[width=0.6\textwidth]{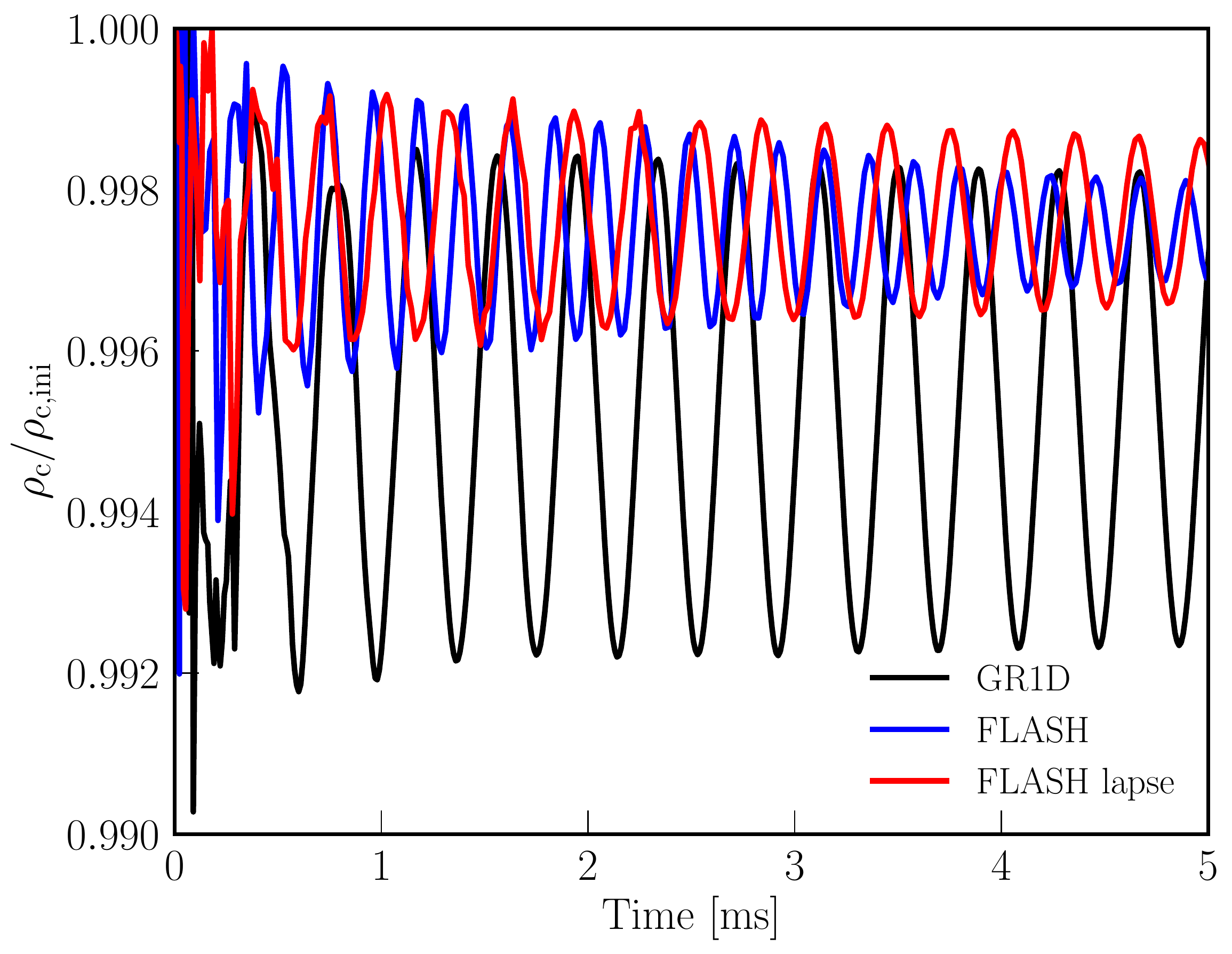}
	\caption{Time evolution of the central density of the stable compact stars in the simulations using GR1D (black), FLASH (blue) and FLASH with the lapse function (red). $\rho_{\rm c,ini}$ is the initial central density. \label{fig:gr}}
\end{figure}

To test the performance of the GR approximations, we simulated the oscillations of a compact star constructed using the Case A potential and hybrid EOS. We choose an initial central density of $1.5\times10^{15}$ g~cm$^{-3}$ to mimic the protocompact star (PCS) after the second bounce in our CCSN simulations. For reference, we also performed a simulation with the GR1D code using a fully relativistic TOV star with the same central density as the initial conditions. The initial conditions are different for the FLASH and GR1D simulations because the fully relativistic TOV solution is not a stable configuration for the FLASH simulations (also see the Appendix A of \cite{2018ApJ...854...63O}). The results of the central density evolution for 5~ms are shown in Fig.~\ref{fig:gr}. The small-amplitude oscillations originate from the imperfect mapping of the initial conditions to the computational grids of the hydrodynamic simulations. This mapping leads to slightly different equilibrium states (central densities) in different codes. The frequencies of the PCS oscillations are $\sim2500$ Hz, $4500$ Hz and $3300$ Hz for the simulations using GR1D, FLASH and FLASH with the inclusion of the lapse function. This test shows that our implementation of the lapse function can approximate the GR time-dilation effect to some extent.

For the CCSN simulations in the main text, we find that the lapse function reduces the gravitational-wave (GW) frequency significantly, especially after the second collapse when the PCS is extremely compact. For example, the frequency of the burst around the second collapse is $\sim2500-4000$ Hz ($\sim4000-5000$ Hz) for the simulation with (without) the lapse function. We expect full GR simulations will further reduce the GW frequency. 

\section{Test of potential numerical artifacts}
\begin{figure}[t]
	\renewcommand{\thefigure}{S\arabic{figure}}
	\includegraphics[width=0.6\textwidth]{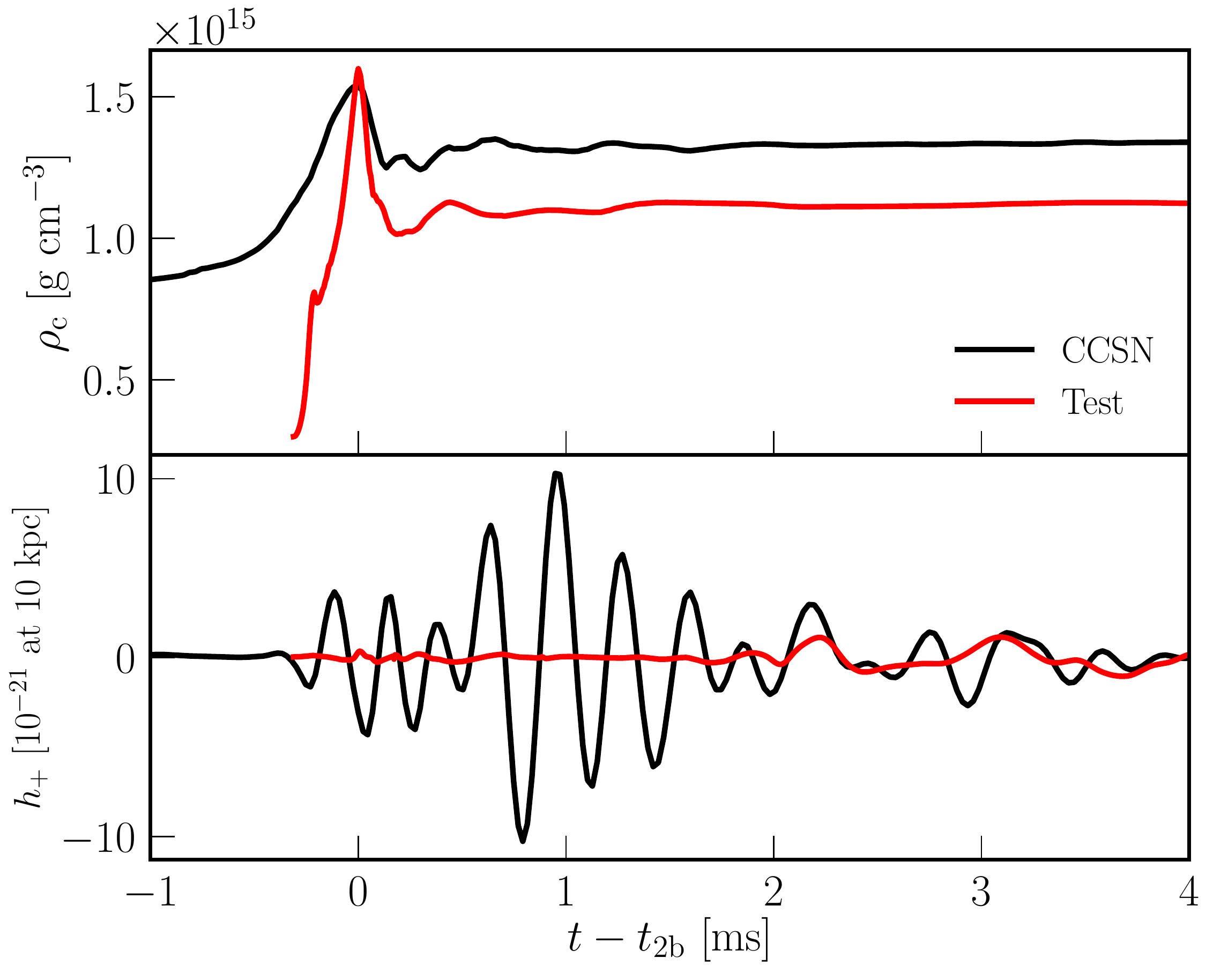}
	\caption{Central density evolution (upper panel) and GW waveform (lower panel) during the PT-induced collapse and postbounce phase for the CCSN (black) and test (red) simulations. \label{fig:test}}
\end{figure}

We perform a test simulation to evaluate the contribution of potential numerical artifacts from the computation domain to the GW signal during the PT-induced collapse and bounce. The test simulation starts from a spherically-symmetric compact star constructed using the STOS EOS, while the hybrid EOS is used for the 2D hydrodynamic simulation. The sudden reduction of pressure results in the collapse of the compact star. We compare the results of this test simulation to those of the PT-induced collapse in the CCSN simulation in Fig.~\ref{fig:test}. In the test simulation, the amplitude of the GW strain remains less than $0.2\times10^{-21}$ until 2 ms after bounce, which is at least an order of magnitude smaller than that of the burst in the CCSN simulation. This indicates that the GW burst in the CCSN simulation results from the asphericity already developed during the episode between the first bounce and the second collapse, but not artifacts from the dynamical collapse itself. 

\section{Resolution dependence}

\begin{figure*}[ht]
	\renewcommand{\thefigure}{S\arabic{figure}}
	\centering
	\includegraphics[width=0.48\textwidth]{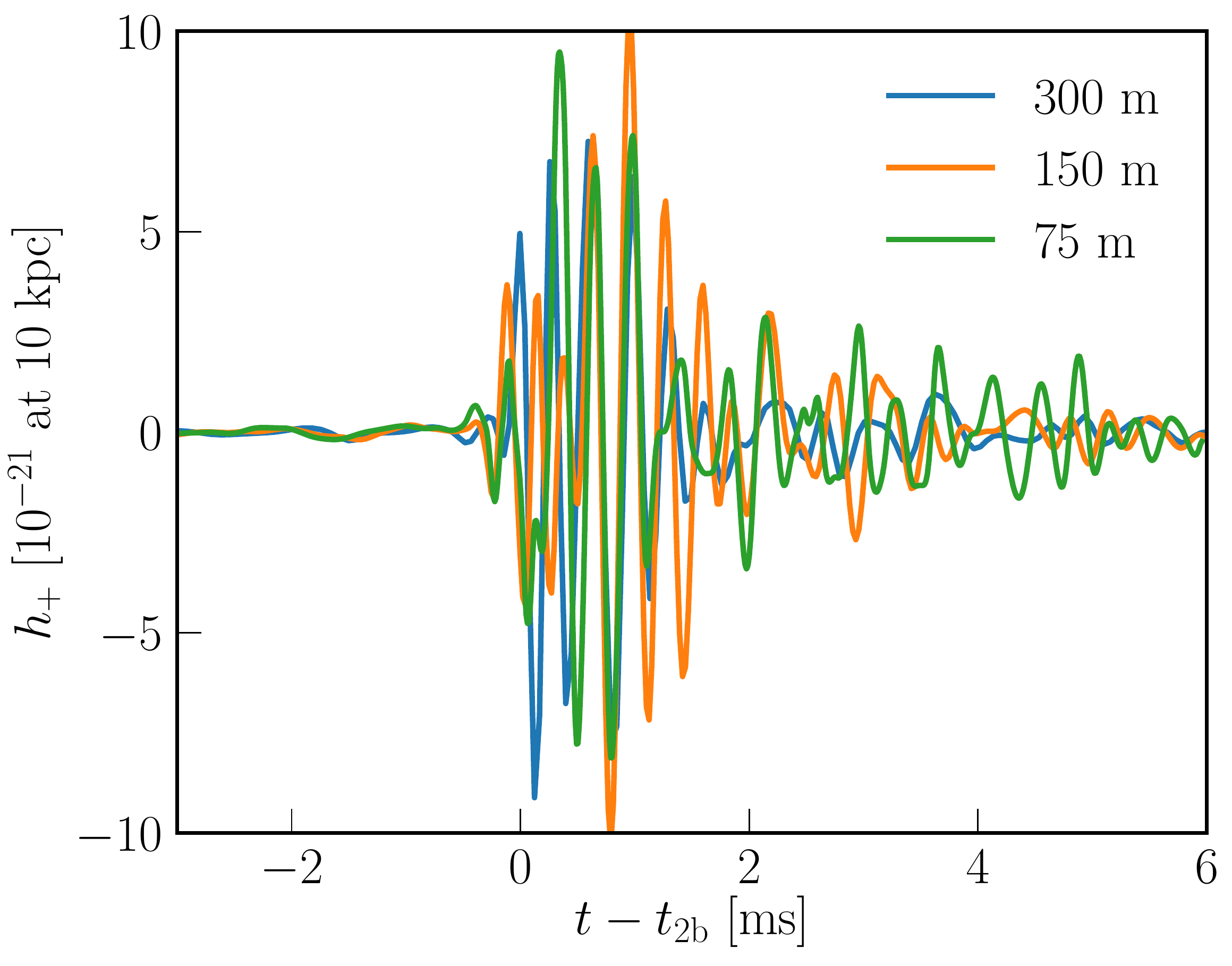}
	\includegraphics[width=0.48\textwidth]{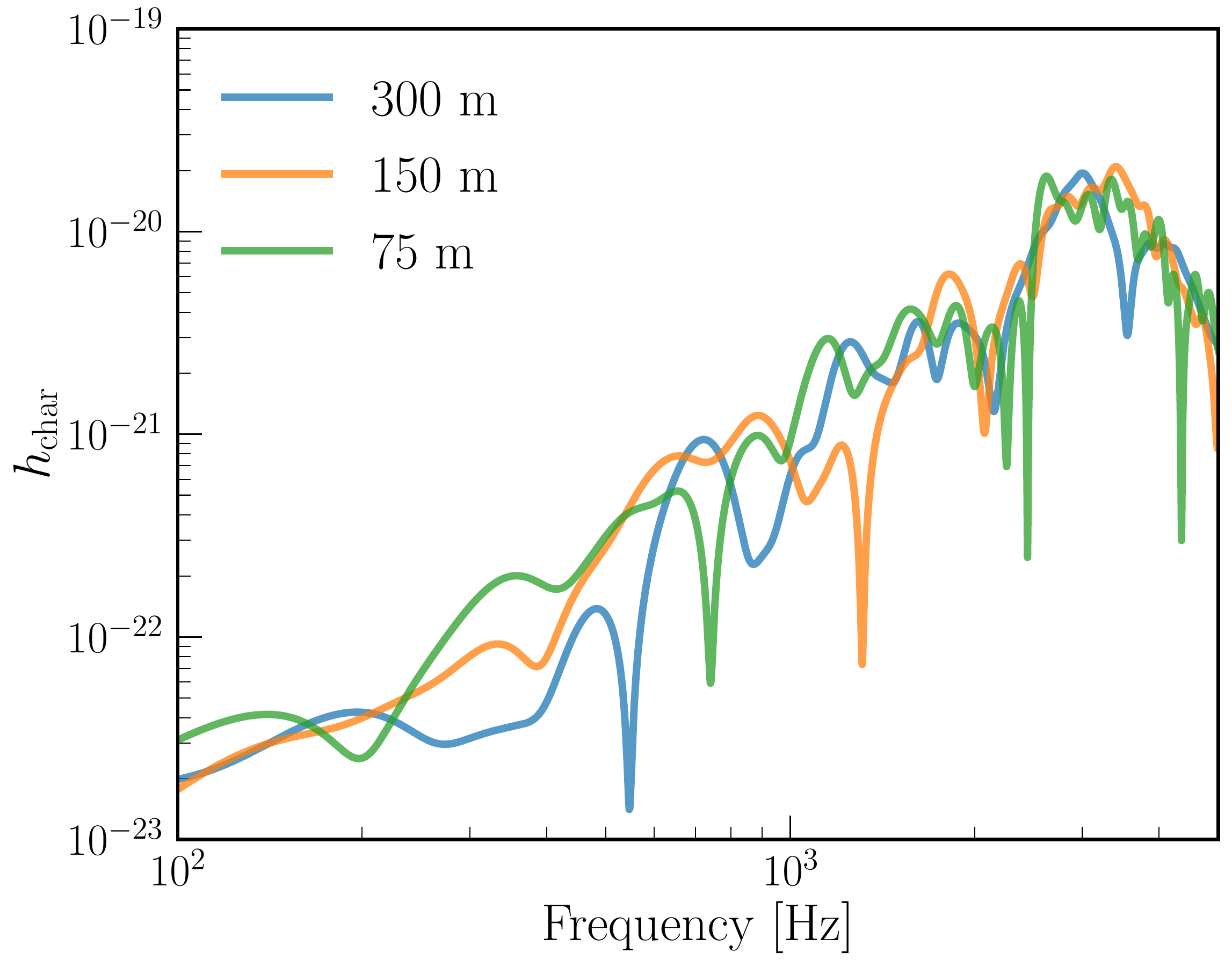}
	\caption{Waveforms (left panel) and characteristic strain spectra (right panel) for the GW burst around the second collapse, extracted from simulations with different resolutions. The legends indicate the finest resolution employed. In the left panel, $t_{\rm 2b}$ refers to the time when the central density reaches its maximum. \label{figa:gw1}}
\end{figure*}

We perform a set of simulations with different resolutions, for the episode of the second collapse and bounce in the main text. The simulations start from $\sim10$ ms before the second collapse and the finest resolutions are 300 m, 150 m and 75 m, respectively. We find the PCS structure after the second bounce agrees well for the different resolutions. In Fig.~\ref{figa:gw1}, we plot the GW waveform and spectrum for the loud burst around the second collapse. Although the GW signals do not match exactly in phase, the amplitude and frequency of the burst agrees quantitatively well for the different resolutions. Because the high-resolution simulation has less numerical dissipation, the amplitude of $h_+$ damps more slowly in the high-resolution runs (75~m and 150~m) than in the low-resolution run (300~m) after $t_{\rm 2b}+2$~ms.

\section{Spatial distribution of GW strain}
\begin{figure*}[t]
	\renewcommand{\thefigure}{S\arabic{figure}}
	\centering
	\includegraphics[width=0.48\textwidth]{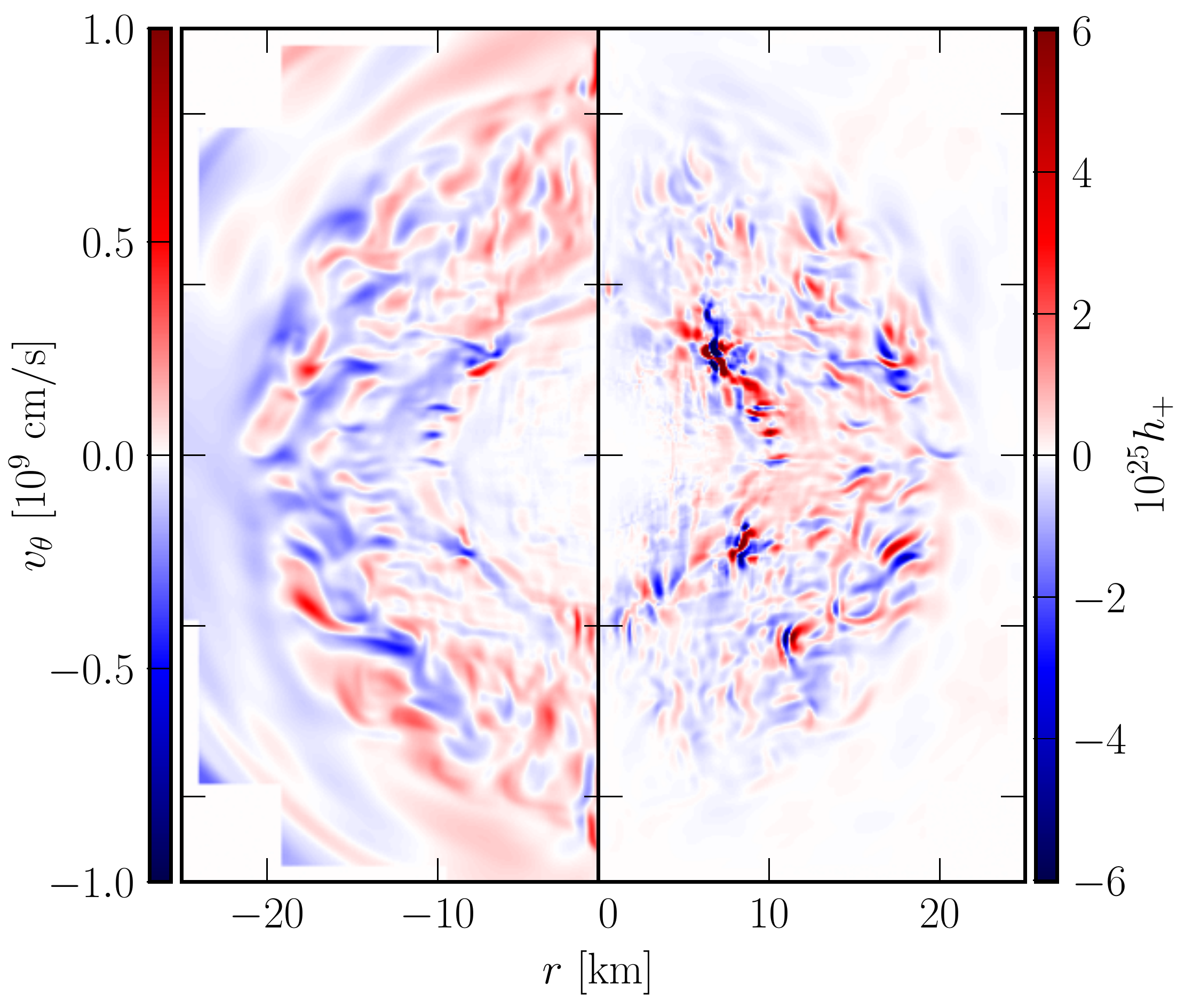}
	\includegraphics[width=0.48\textwidth]{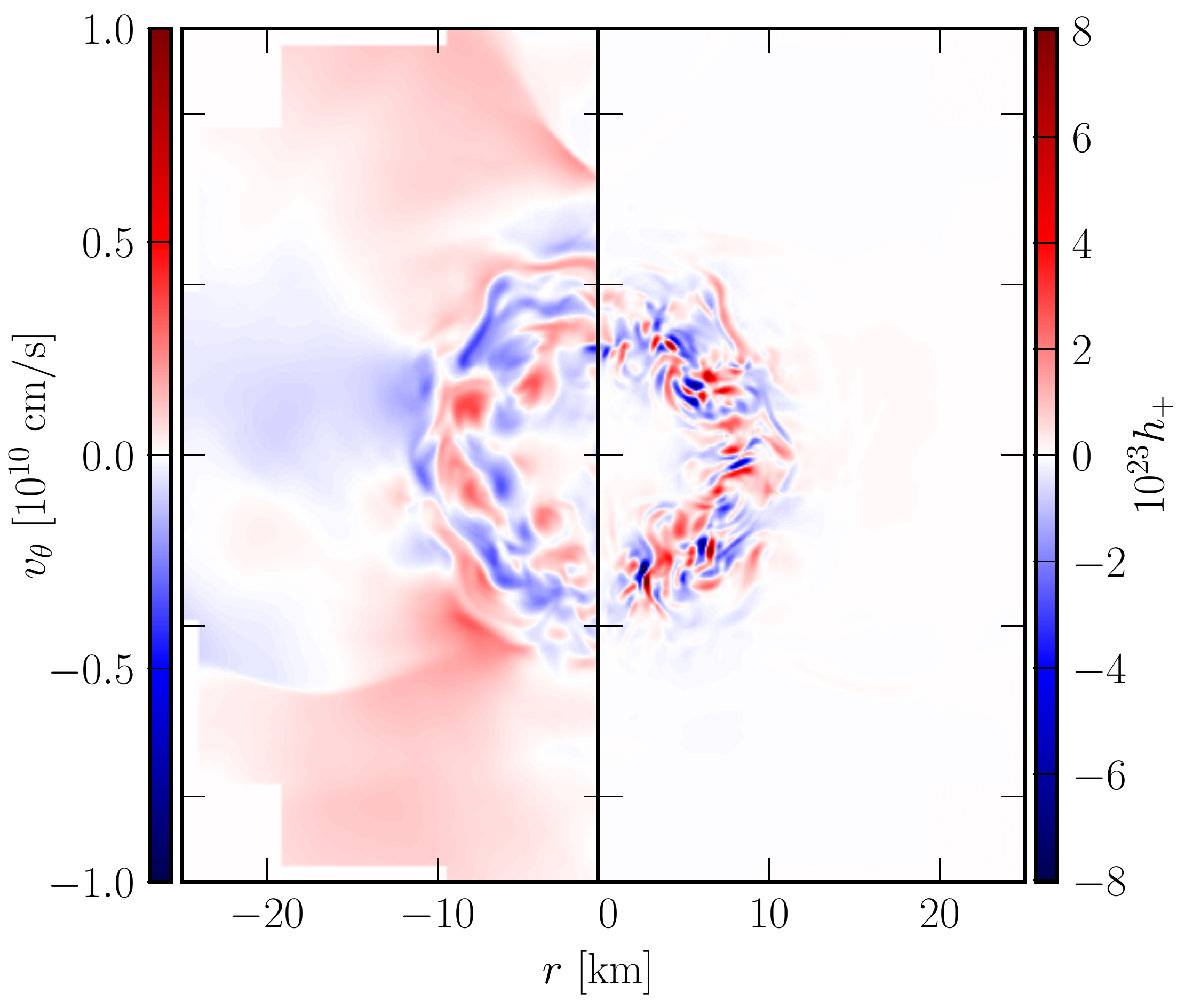}
	\caption{Colormaps for the spatial distribution of the tangential velocity ($v_\theta$, left half) and GW strain ($h_+$, right half) at 2~ms before (left) and 0.5 ms after (right) the second bounce. \label{fig:spatial_h}}
\end{figure*}

Fig.~\ref{fig:spatial_h} shows the spatial distribution of the tangential velocity ($v_\theta$, left half) and GW strain (right half):
\begin{equation}
	h_+ = \frac{2G}{D c^4}\frac{d}{dt} \Big[dV r\rho \Big( v_r P_2(\cos\theta)+\frac{1}{2}v_\theta\frac{\partial}{\partial \theta}P_2(\cos\theta)\Big) \Big], 
\end{equation}
at 2~ms before (left) and 0.5 ms after (right) the second bounce. Here we assume the distance from the source is $D=10~\rm kpc$. The results are from the simulation with a finest resolution of 150 m. The distribution of $v_\theta$ roughly shows the convective regions inside the PCS, which changes from $\sim10-20$~km to $\sim5-10$ km. The distribution of $h_+$ is correlated to that of $v_\theta$, which suggests the connection between the GW emission and convective motions inside the PCS.  The contribution to $h_+$ from the regions outside 15~km is generally less than $\sim10\%$ during the burst around the second collapse.

\section{Spectrogram for the STOS EOS}
\begin{figure}[t]
	\renewcommand{\thefigure}{S\arabic{figure}}
	\includegraphics[width=0.6\textwidth]{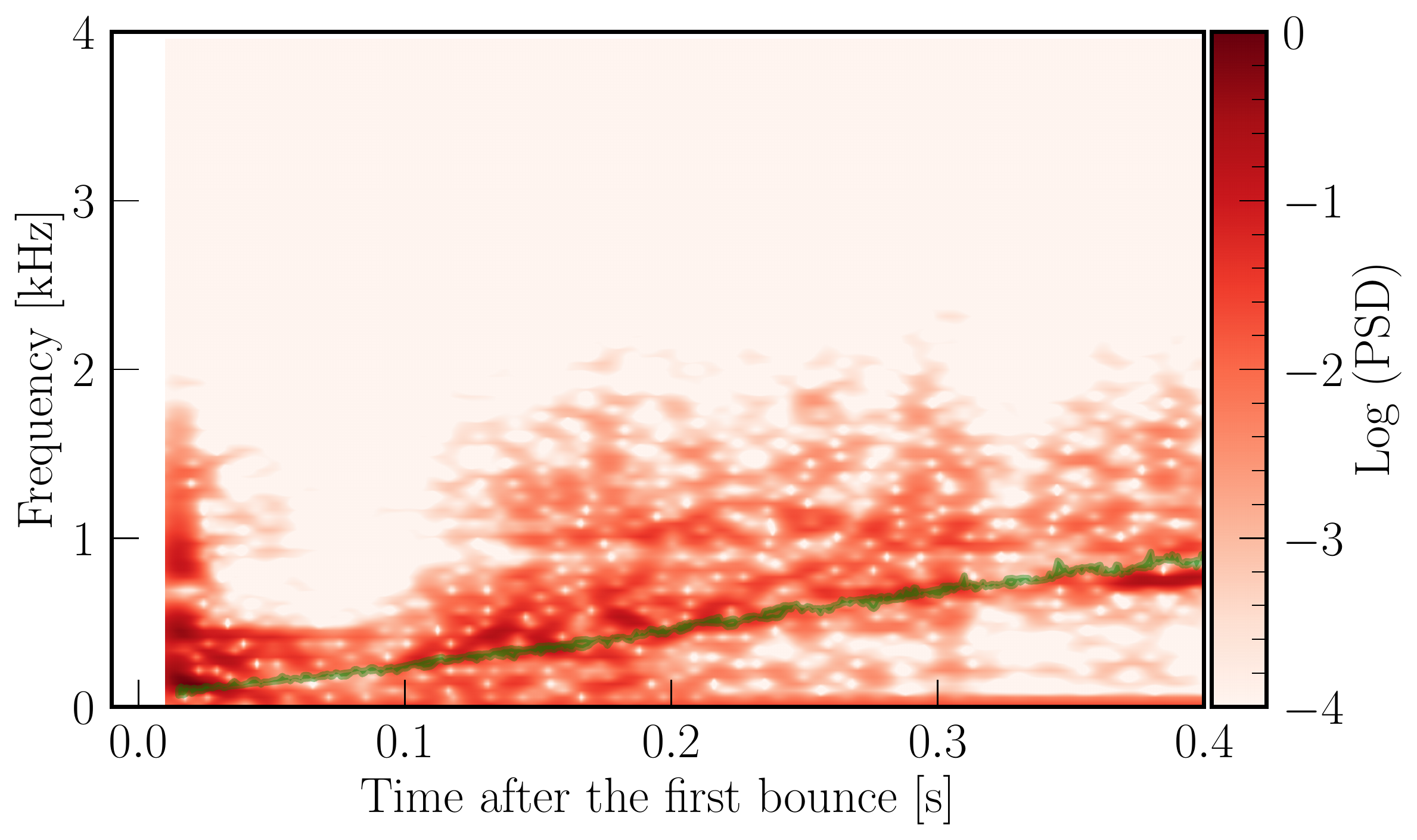}
	\caption{Colormap of the time-dependent power spectral density (PSD) for the GW waveform extracted from the simulation using the STOS EOS. The green-filled bands track the evolution of the Brunt-V\"ais\"al\"a frequencies at densities between $10^{11}$ and $10^{12}$ g cm$^{-3}$. \label{figa:gw2}}
\end{figure}

Fig.~\ref{figa:gw2} shows the spectrogram of the GW signal extracted from the simulation using the STOS EOS. This spectrogram is similar to those of CCSN simulations using a hadronic EOS in the literature, e.g.~\cite{2018ApJ...857...13P}. The green bands shows the time-dependent Brunt-V\"ais\"al\"a frequency ($f_{\rm BV}$) at densities between $10^{11}$ and $10^{12}$~g~cm $^{-3}$, where $f_{\rm BV}$ is independent of the specific position. We estimate $f_{\rm BV}$ by the Newtonian Brunt-V\"ais\"al\"a frequency multiplied by the lapse function:
\begin{equation}
	f_{\rm BV} = \alpha \sqrt{\frac{1}{\rho}\frac{\partial \Phi}{\partial r}\Bigg(\frac{1}{c_s^2} \frac{\partial P}{\partial r} - \frac{\partial \rho}{\partial r}\Bigg)},
\end{equation}
where $c_s$ is the speed of sound and others are the same as in Eq.~\ref{eq:hydro}.
The evolution of the peak GW frequency roughly follows that of $f_{\rm BV}$ for the 400~ms after bounce.

\end{document}